**Morphological and mineralogical characterization of the Freundlich-Sharonov Basin: Implications for the lunar farside magmatism**

Tvisha Kapadia [a, b], Neha Panwar [a], Neeraj Srivastava [a*], Rishitosh Sinha [a], Megha Bhatt [a], Anil Bhardwaj [a]

[a] Planetary Remote Sensing Section, Planetary Sciences Division, Physical Research Laboratory, Ahmedabad, India

[b] Discipline of Earth Sciences, Indian Institute of Technology, Gandhinagar, India

* Corresponding author: Neeraj Srivastava, sneeraj@prl.res.in

**Abstract**

Volcanism on the Moon is highly asymmetric, with the nearside having extensive mare emplacements and the farside exhibiting only sparse, localized volcanism. The Freundlich-Sharonov Basin (FS Basin), a pre-Nectarian/Nectarian impact basin located on the central farside of the Moon (18.35ºN, 175.2ºE), hosts a limited volume of spatially restricted and isolated volcanic patches, which provide an ideal geological setting to investigate controls on the lunar magmatism in a thick, KREEP-poor farside crust. In this study, detailed geological characterization of the FS Basin is provided for the first time, incorporating insights from morphological, chronological, and compositional analyses using high-resolution datasets from missions such as Chandrayaan-1 (Moon Mineralogy Mapper), Kaguya, and Lunar Reconnaissance Orbiter. The results of this study reveal a previously undetected ~192 km-diameter inner depression ring, along with spatially aligned subsurface magmatic intrusions, and exposures of PAN and orthopyroxenes along the basin rings. Compositional analyses indicate that the FS Basin experienced at least two volcanic eruptions dominated by exceptionally high-alumina basalts (~16 – 24 wt% $Al_2O_3$) around ~3.4 Ga and ~2.1 Ga. This is the highest reported alumina content in the lunar maria, also expanding the period of known high- alumina volcanism on the Moon to the late phase. We propose that the limited extent of volcanic eruptions in the FS Basin is a result of the combined effects of composition-dependent magma buoyancy, reduced magma production, and impact-modified crustal structures determining the magma extrusion sites.

## 1. Introduction

Volcanism on the Moon is primarily manifested in the form of mare basalts, which cover ~15.1% of the lunar surface on the nearside compared to only ~1.2% on the farside (Nelson et al., 2014). Mare basalts exhibit significant compositional diversity and are mainly concentrated within and adjacent to the large circular impact structures. These large-scale impact events result in extreme reworking of the crustal structure, which, in turn, influences subsequent magmatic activities and the regional geomorphology. The Freundlich-Sharonov Basin (FS Basin) (18.35ºN, 175.2ºE), ~600 km in diameter, is a pre-Nectarian/Nectarian (pN/N, age: $4.14^{+0.019}_{-0.023}$) basin located on the central farside of the Moon (Orgel et al., 2018). Unlike many lunar impact basins that are completely filled with the mare emplacements, the FS Basin exhibits only a limited distribution of the dark-toned volcanic materials confined within the central portion of the basin. These volcanic materials are mainly hosted within the elongated crater Buys-Ballot and a central pond Lacus Luxuriae, along with sparse distribution within the Anderson E, Anderson F, and Buys-Ballot Q craters (Fig. 1a).

Several studies have previously investigated these dark-toned materials using a variety of datasets from different missions like Clementine, Lunar Reconnaissance Orbiter (LRO), and Chandrayaan-1 (e.g., Besse et al., 2014; Gustafson et al., 2012; Schultz et al., 1998). Schultz et al. (1998) interpreted the Buys-Ballot and Lacus Luxuriae as Fe-enriched friction melts – residual material from a mafic-rich impactor. In support of this claim, the study quoted the isolated nature of the dark-toned features from known volcanic activity and the oblique nature of the Buys-Ballot impact, which could not have produced deep conduits for magma transport due to the shallow disruption levels. Gustafson et al. (2012) suggested that these isolated, dark-toned materials were potential localized pyroclastic deposits and attributed their presence to crustal thinning related to the formation of the FS Basin. Besse et al. (2014) concluded these emplacements to be of volcanic origin, associated with mare volcanism instead of their pyroclastic nature. Thus, there is no clear consensus on the nature of origin and mode of emplacement of these dark-toned materials.

In this study, we conducted a detailed geomorphological and mineralogical analysis of the FS Basin using high-resolution remote sensing datasets. The results provide new insights into the origin, distribution, and chronology of volcanic emplacements within the FS Basin, which help constrain its magmatic and geological evolutionary history. Understanding the magmatism within the FS Basin offers broader implications for identifying the key controls on magmatism in this region, which will be valuable for understanding magmatism in other similar large-scale impact basins.

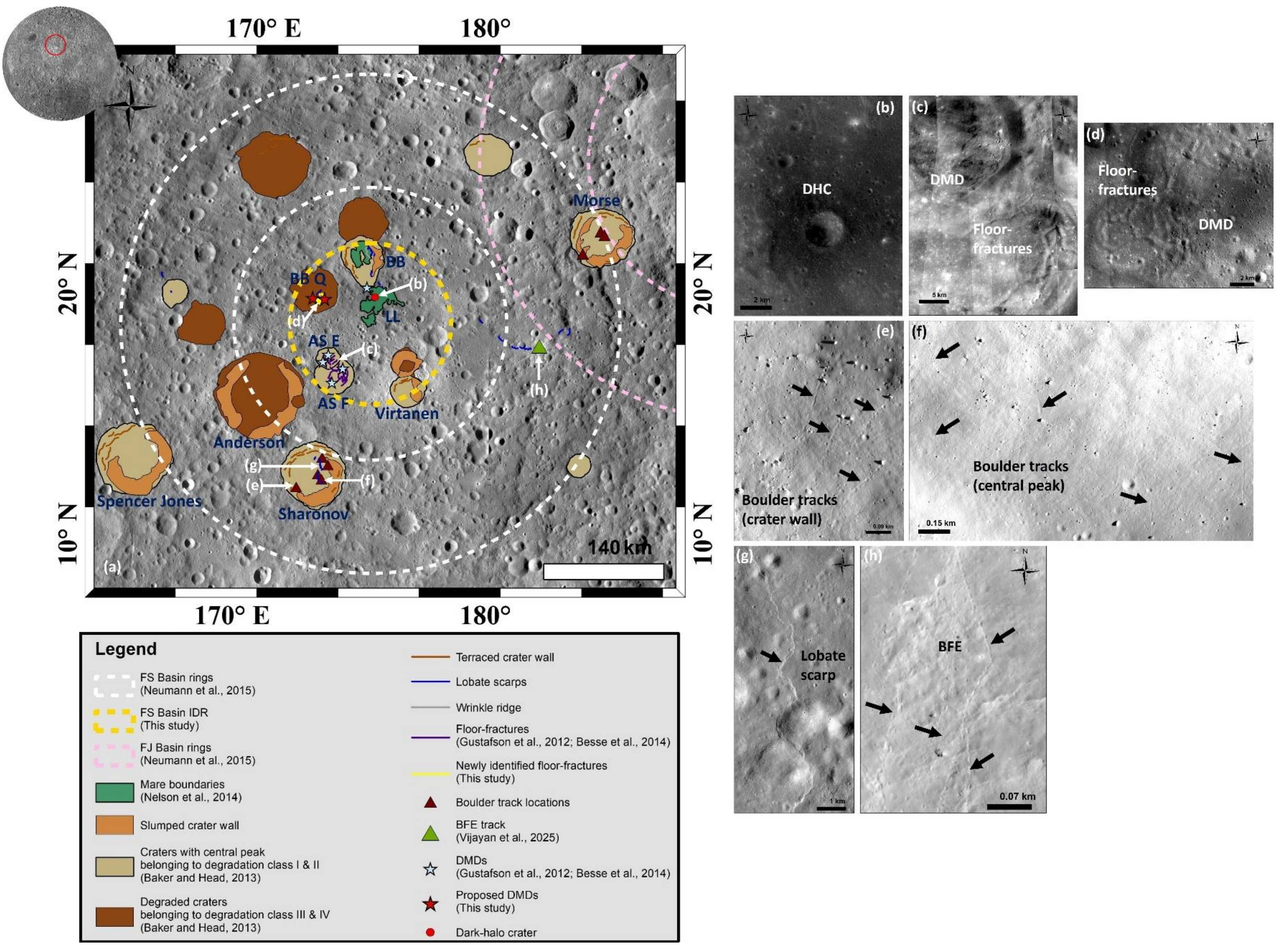

Fig. 1 - (a) Geomorphological map of the Freundlich-Sharonov Basin using the LRO Wide Angle Camera (WAC) mosaic (resolution 100m/pixel). The location of the study area on the Moon is depicted by the red circle in the top left WAC mosaic of the lunar farside. (b-h) High-resolution (0.5 m/pixel) LRO Narrow Angle Camera (NAC) images of the locations marked with white arrows in (a). Here, FS Basin: Freundlich-Sharonov Basin, FJ Basin: Fitzgerald-Jackson Basin, LL: Lacus Luxuriae, BB: Buys-Ballot, BB Q: Buys-Ballot Q, AS E: Anderson E, AS F: Anderson F, BFE: Boulder Fall Ejecta, DMD: Dark Mantle Deposit, and DHC: Dark Halo Crater.

## 2. Datasets and methods

We have carried out a detailed morphological, topographical, compositional, and chronological investigation of the FS Basin using a variety of datasets described in subsequent subsections.

### 2.1 Morphology, topography, and morphometry

The high-resolution images from the LRO Wide Angle Camera (WAC) (resolution: 100 m/pixel) (Speyerer et al., 2011) and Narrow Angle Camera (NAC) (resolution: 0.5 m/pixel) (Robinson et al., 2010) have been used to identify geological features present in the FS Basin and conduct a detailed morphological analysis of the region. The SLDEM2015, a digital elevation model generated by combining Lunar Orbiter Laser Altimeter (LOLA) elevation data and Selenological and Engineering Explorer (SELENE) Terrain Camera (TC) data (Barker et al., 2016) has been used to study the topographic variations within the FS Basin and calculate the essential morphometric parameters. The SLDEM2015 data provides a spatial resolution of 60 m/pixel with a vertical accuracy of ~3-4 m.

The morphometric parameters used in this study are listed in Table 1 and were calculated using topographic profiles generated from SLDEM2015 data. The estimated values are then compared with those predicted from the crater scaling laws, which are expressed in the form of $y = aD^b$, where y is the desired morphometric parameter, D is the crater diameter measured as the distance between rim-to-rim crest, and a and b are constants (Hale & Grieve, 1982; Pike, 1977).

### 2.2 Compositional characterization

We have used the hyperspectral data from the Moon Mineralogy Mapper ($M^3$) onboard Chandrayaan-1, along with the SELENE (Kaguya) Multiband Imager (MI) derived FeO wt%, $TiO_2$ wt%, and $Al_2O_3$ wt% maps, to carry out a detailed mineralogical investigation of the FS

Basin. The compositional maps derived from MI data offer a spatial resolution of 59 m/pixel. The FeO, $TiO_2$, and $Al_2O_3$ maps provide an accuracy of 2 wt%, 0.16 wt%, and 0.09 wt%, respectively (Lemelin et al., 2016; Ohtake et al., 2013; Zhang et al., 2023). The $M^3$ operated in a spectral range of 405-3000 nm and provides a spectral and spatial resolution of 20-40 nm and 140-280 m/pixel, respectively (Green et al., 2011; Pieters et al., 2009). We have utilized radiometrically, photometrically, and thermally corrected (Besse et al., 2013; Boardman et al., 2011; Clark et al., 2011; Green et al., 2011) $M^3$ level 2 reflectance data from the PDS archive (https://pds-imaging.jpl.nasa.gov/volumes/m3.html) to obtain spectral signatures across the study area. The vital step in identifying the mineralogical variations in the region is to carry out a detailed analysis of band parameters (band center, band depth, and band area) of the collected reflectance spectra. In order to estimate these band parameters, the continuum removal of the obtained reflectance spectra was carried out using ENVI software's continuum removal tool following the methodology given by Clark et al. (1987). The data has been clipped at 2538 nm to avoid the lunar surface emission component.

The mineralogy of the study area has been interpreted based on the analysis of the diagnostic absorption features centered near 1000 nm (Band I) and 2000 nm (Band II) arising due to the presence of $Fe^{2+}$ in the crystallographic sites (Burns, 1981, 1993; Cloutis et al., 2008). The Band I Center (BC I) and Band II Center (BC II) were estimated using the method described by Gustafson et al. (2020), where BC I and BC II correspond to the band minima of the biquadratic polynomial fitted within the ±200 nm wavelength range of the respective band minima. The relative positions of BC I and BC II obtained from the continuum-removed spectra have been used for the identification of orthopyroxenes (opx) or Low Ca-pyroxenes (LCP) and clinopyroxenes (cpx) or High Ca-pyroxenes (HCP). The LCP have BC I and BC II positions estimated at ~900-940 nm and ~1800-2100 nm, respectively, whereas the HCP have BC I and BC II positions at ~960-1050 nm and ~2100-2400 nm (Adams, 1974; Cloutis & Gaffey, 1991a, 1991b). The band areas were calculated by estimating the areas enclosed by the fitted continua and the Band I and/or Band II of a continuum-removed spectrum. The anchor points for Band I were set at 699 nm and 1578 nm, and the Band II were set at 1578 nm and 2538 nm for the band area estimation. The BC I values have been plotted against BC II to classify the pyroxene spectra into opx and cpx (Adams, 1974). In order to determine the mafic compositions in an olivine-pyroxene mixture, the BC I values have been plotted against the Band Area Ratio (BAR) (Band II Area / Band I Area) (Zhang & Cloutis, 2021a, 2021b).

To highlight the compositional variability present in the study area, we generated an Integrated Band Depth (IBD) based false colour composite (FCC). The IBD refers to the integration of all the band depths over a specified wavelength range of the continuum-removed spectrum. The IBD I (Band I) and IBD II (Band II) values were estimated using the methodology given by Cheek et al. (2011) for the $M^3$ data:

$$\text{IBD I} = \sum_{n=0}^{26} 1 - \frac{R(789+20n)}{Rc(789+20n)}$$

$$\text{IBD II} = \sum_{n=0}^{21} 1 - \frac{R(1658+40n)}{Rc(1658+40n)}$$

The IBD-based FCC was prepared by assigning red to the IBD I image, green to the IBD II image, and blue to the 1489 nm reflectance image. The resultant FCC represents the mafic lithologies, where the $Fe^{2+}$ absorption is stronger, in yellow, green, and orange colours. On the other hand, the lithologies having weaker absorption, either due to the abundance of feldspathic material and/or higher rate of space weathering, yield blue colour in the FCC.

### 2.3 Chronological analysis

To understand the evolutionary sequence of impact and magmatic events within the FS Basin, we carried out a detailed chronological study of the selected targets in and around the central depression of the FS Basin. We used the Crater Size Frequency Distribution (CSFD) technique (Neukum et al., 2001) to determine the ages of the Fe-rich mare basalts in Lacus

Luxuriae and Buys-Ballot, and the Fe-poor Virtanen crater. To perform crater-counting of the selected targets, we have used high-resolution images from Lunar Reconnaissance Orbiter (LRO) NAC (0.5 m/pixel resolution) (Robinson et al., 2010). While performing crater counting, we have carefully selected the topographically uniform units and excluded secondary crater chains and clusters (Michael & Neukum, 2010). To minimize the effect of the smaller obliterated impact craters on the CSFD, we have applied a cumulative resurfacing correction. The software used for crater-counting is ArcGIS, along with the CraterTools (v2.1) extension. The Absolute Model Ages (AMAs) were determined using Craterstats2.

## 3. Geological setting of the Freundlich-Sharonov Basin

The Freundlich-Sharonov Basin, ~600 km in diameter, is a pre-Nectarian/Nectarian (pN/N) basin located on the central farside of the Moon (18.35ºN, 175.2ºE) (Fassett et al., 2012; Orgel et al., 2018). The Freundlich-Sharonov Basin (FS Basin) is highly degraded by the superposed impacts (Fig. 1a). The existence of this nearly obliterated basin was first suggested by photogeological interpretations by Stuart-Alexander (1978) and Wilhelms et al. (1987), and later confirmed by Clementine altimetry data (Spudis et al., 1994). Using the Gravity Recovery and Interior Laboratory (GRAIL) mission data, Neumann et al. (2015) identified the presence of a peak-ring (diameter: ~318 km) and an outer ring (diameter: ~582 km) associated with the FS Basin. On the east, the outer ring of the FS Basin intersects the outermost ring of the adjacent Fitzgerald-Jackson (FJ) Basin (Fig. 1a), which does not exhibit any topographic or morphological expression (Neumann et al., 2015).

Unlike many lunar impact basins that are completely filled with the mare emplacements, the FS Basin exhibits only a limited distribution of the volcanic materials confined within the central portion of the basin. The central region of the FS Basin hosts a mare patch, Lacus Luxuriae (LL), which exhibits a smooth, low-albedo surface (Fig. 1a). The northern part of the LL has a lower albedo compared to the Southern part (Morota et al., 2011). The Buys-Ballot (BB), located north of the LL, is an elongated complex crater formed by a highly oblique impact (Fig. 1a) (Schultz et al., 1998). The BB crater also has volcanic material emplaced on its floor. Anderson E, Anderson F, and Buys-Ballot Q craters exhibit distinct crater floor morphology (Fig. 1c, d). These craters show the development of floor fractures, and their crater floors are partially covered by the Dark Mantle Deposits (DMDs) (Fig. 1c, d) (Besse et al., 2014; Gustafson et al., 2012). The DMDs are primarily characterized by their low albedo and mantling relationship to the underlying terrain features, and are often associated with explosive volcanic activity (Head, 1974; Heiken et al., 1974; Rosanova et al., 1998), suggesting the possible presence of pyroclastics in this region (Besse et al., 2014; Gustafson et al., 2012).

## 4. Results and discussions

### 4.1 Topographic, Morphologic, and morphometric analysis

The topographic map of the highly degraded FS Basin, generated using SLDEM2015, reveals a discontinuous, elevated topographic ring, locally disrupted by several superposed impact craters, defining an almost circular outline with a diameter of ~192 km around the basin's central depression (Fig. 2a). The topographic profiles along XX' and YY' also show raised elevation at these locations (Fig. 2b, c), which indicates the possible presence of a previously undetected Inner Depression Ring (IDR) composed of fault scarps or discontinuous massifs.

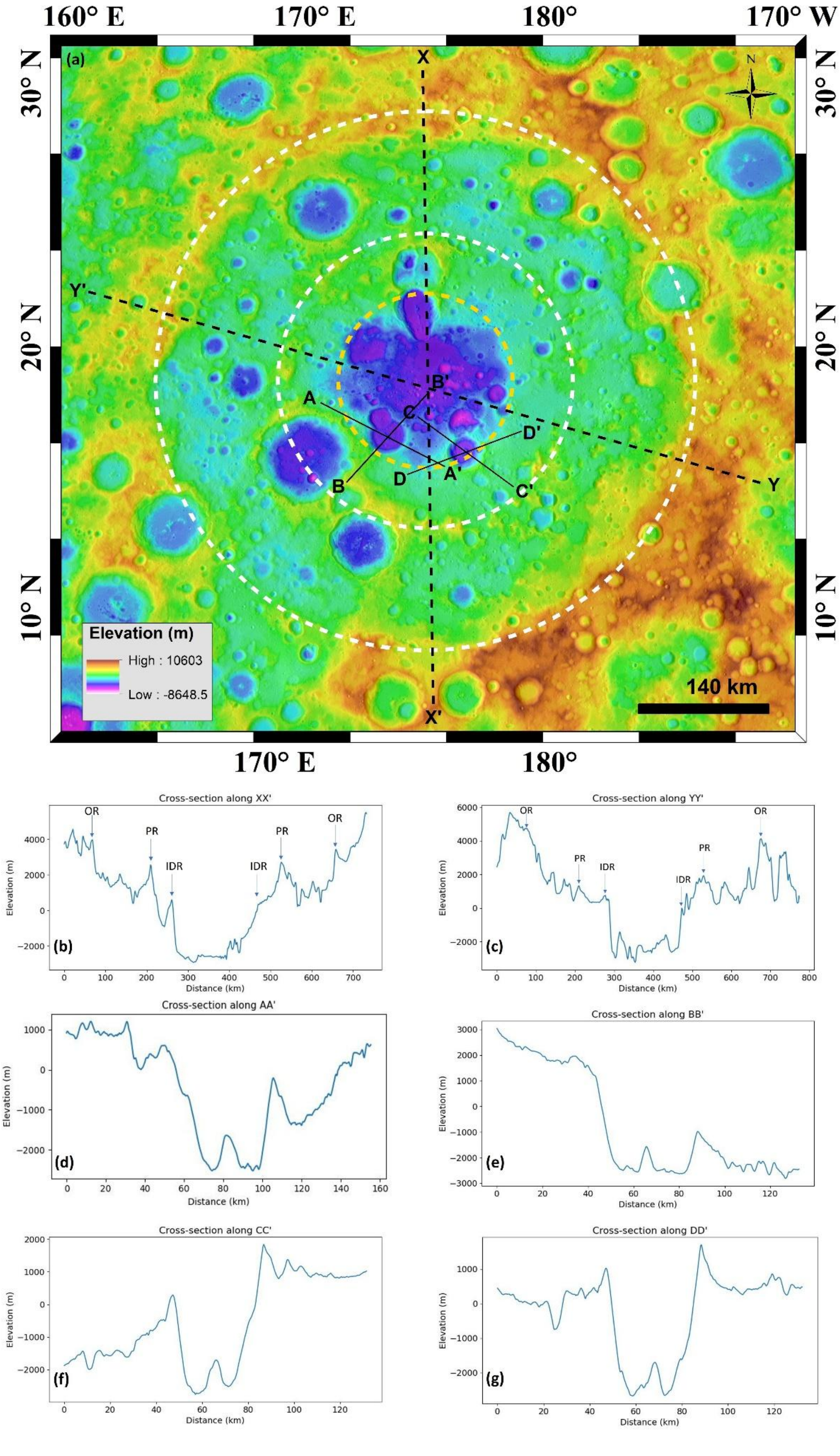
160° E
170° E
180°
170° W
30° N
20° N
10° N
(a)
X
X'
Y
Y'
A
A'
B
B'
C
C'
D
D'
N
Elevation (m)
High : 10603
Low : -8648.5
140 km
Cross-section along XX'
OR
PR
IDR
Elevation (m)
Distance (km)
(b)
Cross-section along YY'
(c)
Cross-section along AA'
(d)
Cross-section along BB'
(e)
Cross-section along CC'
(f)
Cross-section along DD'
(g)

Fig. 2 - (a) A SLDEM2015 based topographic map of the FS Basin overlain on LRO WAC mosaic, showing the transects along which the topographic profiles have been obtained. White dashed circles denote the rings of the FS Basin suggested by Neumann et al. (2015). (b, c) Topographic cross-sections of the FS Basin along XX' and YY' exhibiting raised rings associated with the basin. Here, OR is the outer ring, and PR refers to the peak ring of the basin. The yellow dashed circle in (a) represents the proposed Inner Depression ring (IDR) of the FS Basin. (d-g) Topographic cross-sections of the floor-fractured Anderson F crater (AA' and BB') and Virtanen crater (CC' and DD').

Volcanic emplacements within the FS Basin are confined within the central depression encompassed by the proposed IDR of the basin. The mare emplaced within BB and LL covers a surface area of ~500 km$^2$ and ~1000 km$^2$, respectively. The mare on the floor of the BB crater is superposed by multiple wrinkle ridges, indicating a localized compressional stress regime following the basaltic infilling. The terrain downrange from the BB crater, transitioning into the mare-filled LL, exhibits a hummocky-textured surface. A crater ~2.5 km in diameter located in the central portion of the LL is surrounded by a dark halo having very low albedo. Based on the presence of a dark halo and a distinct mineralogy excavated by this crater, we categorized it as a Dark Halo Crater (DHC) (Fig. 1b).

The Anderson F (Diameter: ~44 km), Anderson E (Diameter: ~27 km), and Buys-Ballot Q (Diameter: ~55 km) craters, located along the proposed IDR of the FS Basin, show distinctive floor-fractured morphology, indicative of either the viscous relaxation (Dombard & Gillis, 2001; Hall et al., 1981; Schultz, 1976) of the crater floor or the presence of the subsurface magmatic intrusions (Jozwiak et al., 2012; Schultz, 1976; Thorey & Michaut, 2014; Wichman & Schultz, 1995). However, for a crater to relax viscously, its diameter should be larger (> 60 km) (Dombard & Gillis, 2001) than that of Anderson F, Anderson E, and Buys-Ballot Q craters. Therefore, the viscous relaxation is an unlikely mechanism for the floor-fractured morphology observed in these craters, and the floor-fracturing is more likely attributed to the magmatic intrusion-driven uplift in this region. The other mechanism that attributes the floor fracturing to the presence of subsurface magmatic intrusion states that upon impact on a surface, a highly fractured brecciated zone forms beneath the crater. When a dike rises towards the surface, at the base of the crater, it encounters this low-density brecciated region and becomes neutrally buoyant (Head & Wilson, 1992), thereby ceasing its vertical movement. The magma then starts to spread laterally, forming a sill beneath the crater that exerts pressure on the crater floor, leading to vertical thickening, a piston-like uplift, and fracturing of the crater floor (Jozwiak et al., 2012; Schultz, 1976).

The Virtanen crater (Diameter: ~40 km), juxtaposed with the Anderson F and Anderson E craters, is nearly identical in size to Anderson F and does not exhibit the floor-fractured morphology. Instead, the Virtanen crater retains a relatively smooth crater floor devoid of any fracturing. This morphological difference is quite remarkable as all four craters – Anderson F, Anderson E, Buys-Ballot Q, and Virtanen are situated in the same geological setting within the central depression of the FS Basin and are expected to have undergone similar morphological evolution. The absence of floor fracturing in the Virtanen crater raises an important question of whether the magmatic intrusions beneath the Anderson F, Anderson E, and Buys-Ballot Q have a limited extent and are spatially localized, or whether they also extend into the region beneath the Virtanen crater. In the latter case, it is possible that the Virtanen has not experienced the same level of intrusion-related modifications as the other craters due to differences in the magnitude or extent of the magmatic intrusion.

To investigate the presence of the subsurface magmatic intrusions in this region, we have carried out a detailed morphometric analysis of the two similarly sized craters – the Anderson F and the Virtanen. Using SLDEM2015 data, we generated topographic profiles by taking cross-sections across AA’, BB’, CC’, and DD’ (Fig. 2d-g) to extract morphometric parameters, including crater diameter, crater depth, rim height, central peak height, and basal area of the central peak (Table 1). The extracted values of the morphometric parameters from multiple topographic cross-sections were averaged for each crater and compared with those predicted from the crater scaling laws as a function of crater diameter (Hale & Grieve, 1982; Pike, 1977).

Table 1 – Morphometric parameters used in the study, their definitions, and formulae *([q]Hale & Grieve, 1982; [p]Pike, 1977)*.

| **Morphometric Parameters** | **Definition** | **Formulae** | **Crater** | | | |
|---|---|---|---|---|---|---|
| | | | **Anderson F** | | **Virtanen** | |
| | | | **Measured value** | **Predicted value** | **Measured value** | **Predicted value** |
| **Diameter** | Rim-to-rim crest distance | | 44.22 km | | 40.43 km | |
| **Crater depth** | Elevation difference between the average crater floor and the rim crest | $1.044*D^{0.301}$[p] | 2.63 km | 3.27 km | 2.51 km | 3.18 km |
| **Rim height** | Height of the rim crest above the pre-existing surface | $0.236*D^{0.399}$[p] | 2.21 km | 1.07 km | 1.26 km | 1.03 km |
| **Central peak height** | Height of the central peak crest from the crater floor | $0.0006*D^{1.97q}$ | 0.83 km | 1.04 km | 0.85 km | 0.87 km |
| **Basal area of the central peak** | Area covered by the base of the central peak | $0.009*D^{2.19q}$ | 143.65 $km^2$ | 36.15 $km^2$ | 83.54 $km^2$ | 29.71 $km^2$ |

It is clear from Table 1 that the measured values using topographic cross-sections differ significantly from those predicted using crater scaling laws. Both craters exhibit shallower depths and reduced central peak heights, along with increased rim heights, compared to those predicted by the crater scaling laws. The observed increase in rim height of both craters further provides additional support, alongside the diameter constraint, that viscous relaxation is not responsible for the floor-fractured morphology, as it typically preserves or reduces the rim height due to the relaxation of topographic relief (Dombard & Gillis, 2001). The observed differences between the measured and predicted values can therefore be attributed to the variability in the geological setting in which they formed and/or to the post-impact modifications resulting from the presence of magmatic intrusions. To further examine the sub-crater magmatic intrusion mechanism, we analysed the local geological setting of the region. Both the craters Anderson F and Virtanen are located along the IDR of the ~600 km Freundlich-Sharonov basin. The impact that has caused the formation of such a huge basin would have led to a large network of fractures beneath this basin, and the subsurface of the central portion of this basin would be highly fractured and brecciated. This is an extremely favourable situation for the propagation of the dike from the magma source to the surface. When this dike encounters a brecciated lens, it achieves neutral buoyancy and stalls. As the exerted pressure on the dike increases, it starts spreading laterally below the crater, resulting in an uplift of the crater floor, which shallows the depth of the crater and increases the rim height above the predicted value. All of these morphometric deviations from the predicted values are observed in both Anderson F and Virtanen craters, indicating that both craters have experienced subsurface magmatic intrusion-related modifications. Thus, the presence of magmatic intrusions beneath the Virtanen crater is clearly evident even though it has not developed a typical floor-fractured morphology. The greater magnitude or lateral extent of the magmatic intrusions beneath Anderson F, Anderson E, and Buys-Ballot Q may have exerted sufficient pressure on the crater floor from beneath, which would have produced sufficient uplift and stress regime that eventually resulted in the development of fractures on the crater floor, whereas the magmatic intrusions beneath Virtanen may not have reached the threshold required to produce the floor-fractured morphology. Additionally, all of these craters have formed along the proposed IDR of the FS Basin. The basin rings are typically associated with deep-reaching faults, which serve as pathways for the ascent of magma (Andrews-Hanna et al., 2018; Johnson et al., 2016; Nahm et al., 2013). The deduced presence of magmatic intrusions beneath the craters situated along the proposed IDR, therefore, gives additional support for the existence of the inner ring structure. The subsurface ring dikes associated with the proposed inner ring scarps may have provided conduits for magma to ascend, form subsurface intrusions, and cause volcanism along the innermost ring of the FS Basin.

The Spencer Jones (~90 km diameter) and Morse (~75 km diameter) craters, situated along the outer ring of the FS Basin, show heavily slumped crater walls and formation of wall terraces (Fig. 1a). In addition to the outer ring of the FS Basin, the proposed inner ring of the adjacent FJ Basin (Neumann et al., 2015) also passes through the Morse crater, making this crater an interesting target for understanding basin-ring tectonic-related modifications. The Anderson (~105 km diameter) and Sharonov (~75 km diameter) craters are located along the peak ring of the FS Basin (Fig. 1a). The Anderson crater is a highly degraded crater with slumped crater walls. The north-eastern and eastern regions of the Sharonov crater wall, located in close proximity to the FS Basin peak ring, exhibit extensive slumping and a distinct terraced morphology. The southern part of the crater wall shows highly slumped material and lacks a distinct terrace formation, whereas the eastern wall shows well-developed terraces (Fig. 1a).

High-resolution LROC NAC images revealed high rock abundance and numerous boulder tracks on the Morse and Sharonov crater walls and central peak region (Fig. 1a, e, f). Previous studies have shown that the survival time of boulder tracks on the lunar surface is very short (~35 Ma) (Ruj et al., 2022; Senthil Kumar et al., 2016). Thus, the preservation of these boulder tracks suggests that recent mass wasting and tectonic activity have occurred in the region. The floor of the Sharonov crater also shows the presence of lobate scarps (Fig. 1g), which also suggest the region had been tectonically active in the recent geological past (Clark et al., 2024; van der Bogert et al., 2018). Additionally, multiple Boulder Fall Ejecta (BFE) tracks are present in a crater proximal to a lobate scarp located near the west of the FS Basin peak-ring (Fig. 1h). The presence of BFE tracks in the close vicinity of the lobate scarp is also interpreted as a marker of mass wasting triggered by recent endogenic tectonic activity within the past few decades to a few thousand years (Vijayan et al., 2025). Therefore, the presence of boulder tracks, lobate scarps, and BFE collectively indicates that the FS Basin region has remained tectonically active until very recent geological time.

**4.2 Compositional investigation and chronological framework**

The study utilized the $M^3$-derived IBD-based FCC (Fig. 3), along with the Kaguya-derived FeO wt% (Fig. 4a), $TiO_2$ wt% (Fig. 4b), and $Al_2O_3$ wt% (Fig. 4c) maps to highlight the compositional diversity within the FS Basin.

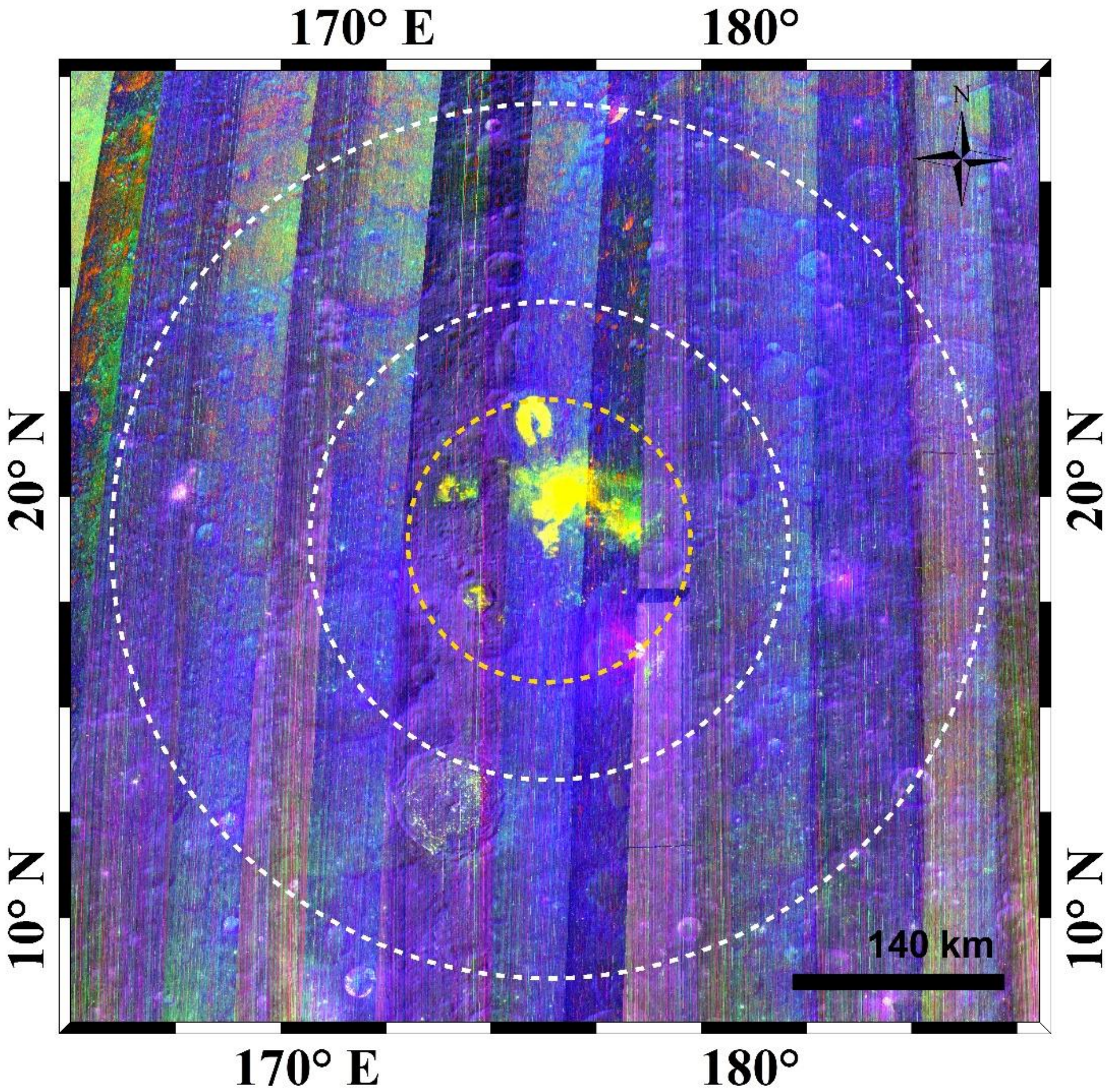

Fig. 3 – $M^3$-derived IBD-based FCC (R: IBD I, G: IBD II, B: Reflectance at 1489 nm) of the Freundlich-Sharonov Basin. Dashed boundaries represent the rings of the FS Basin (Yellow: proposed IDR in this study).

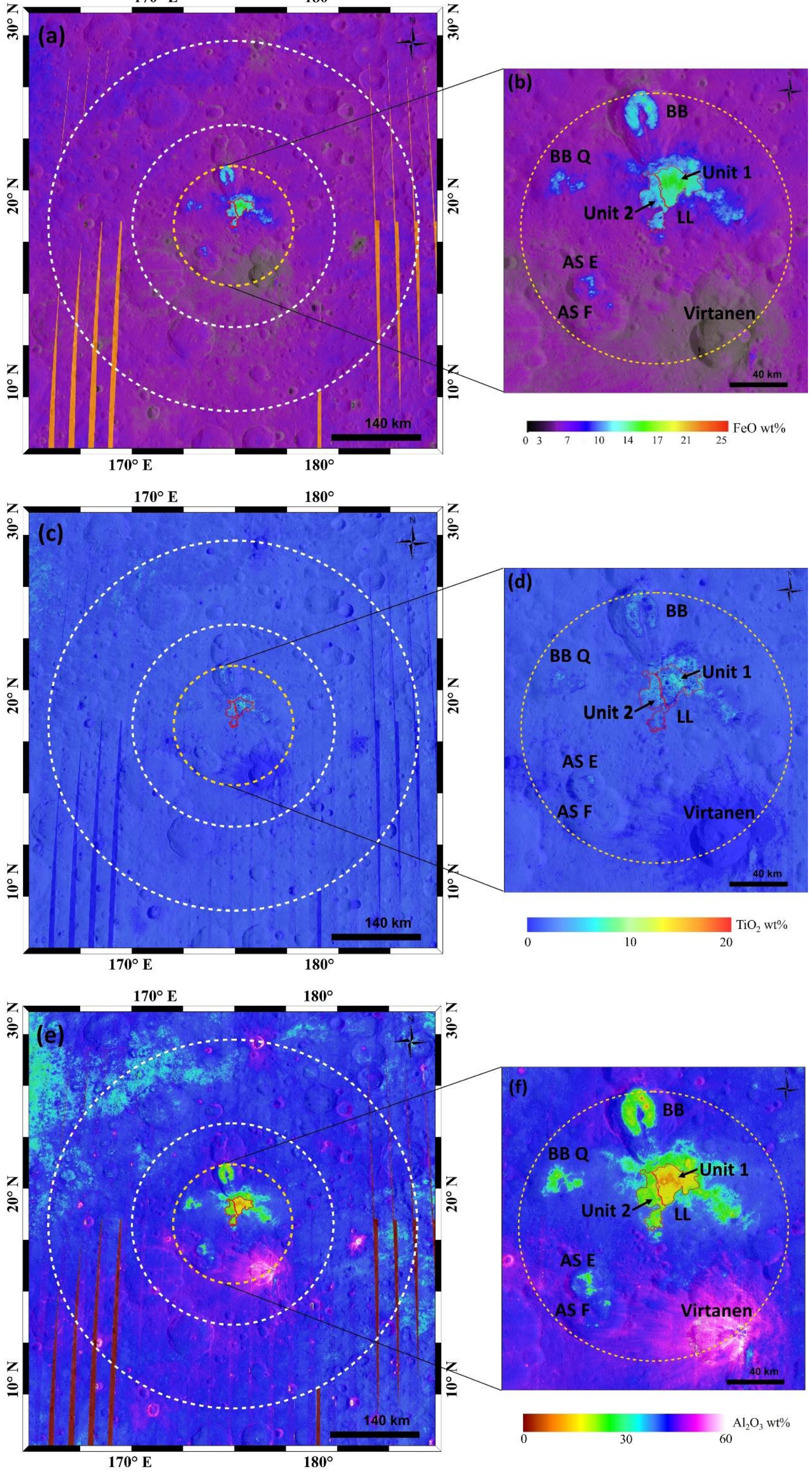

170° E
180°
30° N
20° N
10° N
(a)
140 km
(b)
BB
BB Q
Unit 1
Unit 2
LL
AS E
AS F
Virtanen
40 km
FeO wt%
0 3 7 10 14 17 21 25
(c)
(d)
TiO$_2$ wt%
0 10 20
(e)
(f)
Al$_2$O$_3$ wt%
0 30 60

Fig. 4 – The Kaguya-derived (a) FeO wt%, (c) $TiO_2$ wt%, (e) $Al_2O_3$ wt% maps of the FS Basin overlain on LRO WAC mosaic. (b, d, and f) are zoomed in view of the central region of the FS Basin shown in (a, c, and e), respectively. Dashed white and yellow circles mark the rings of the basin. The area encompassed by the red line represents the two compositional units identified within the Lacus Luxuriae. Here, LL: Lacus Luxuriae, BB: Buys-Ballot, BB Q: Buys-Ballot Q, AS E: Anderson E, AS F: Anderson F.

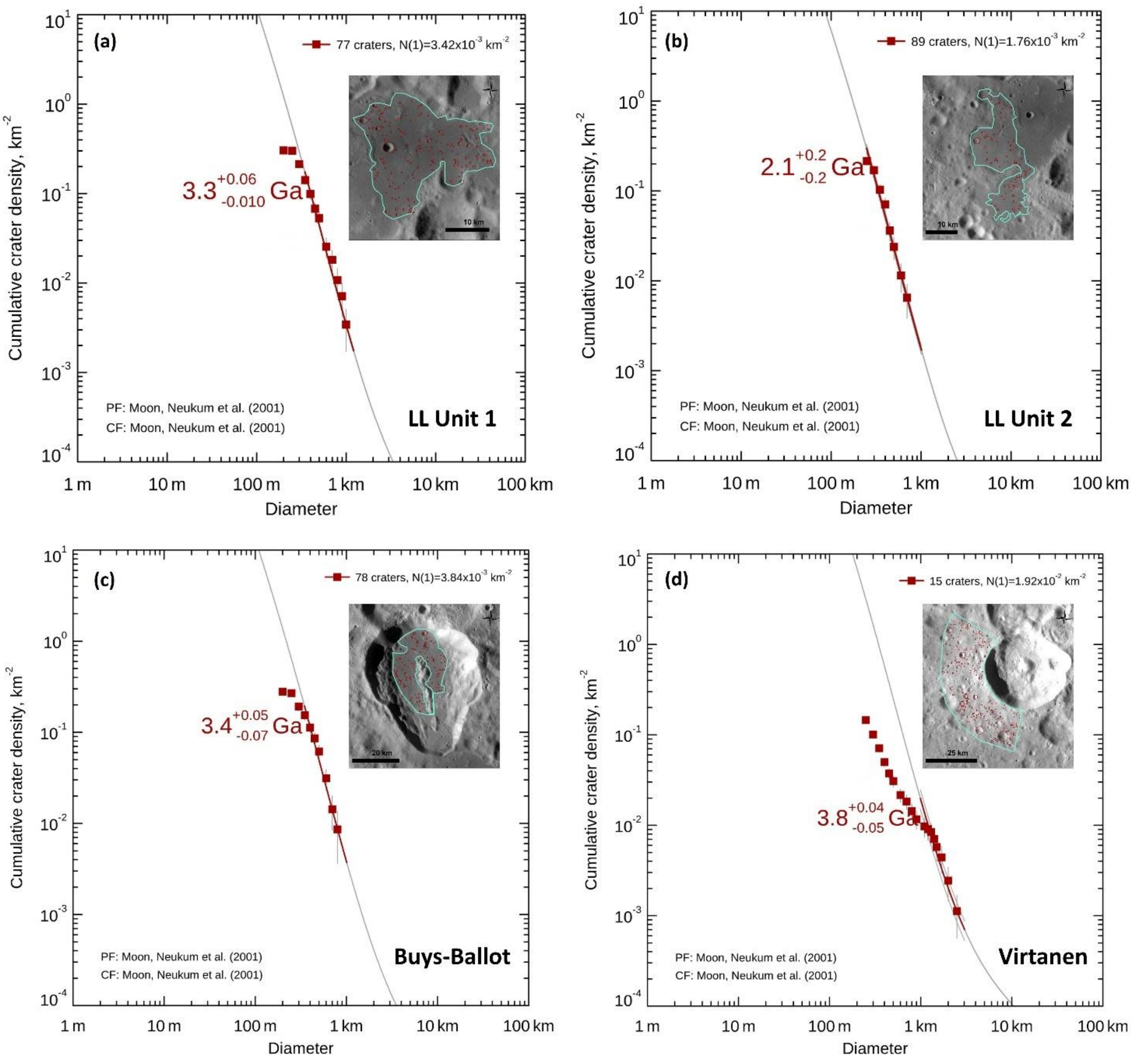

Fig. 5 - The CSFD plots of (a) Lacus Luxuriae Unit 1 (High FeO unit), (b) Lacus Luxuriae Unit 2 (Low FeO unit), (c) Buys-Ballot mare, and (d) Virtanen crater. The LRO WAC images in the top right corners depict the region where crater counting was performed to determine their ages. The cyan line marks the defined area for crater counting, and the red circles are the craters counted.

#### 4.2.1 Lacus Luxuriae and Buys-Ballot

The basalts of LL and BB comprise the central region of the FS Basin, situated within the topographic depression formed as a result of basin-forming impact (Fig. 1a). Based on the variation in the FeO concentrations, we have demarcated two distinct mare units within the LL. The eastern unit (Unit 1) shows higher FeO concentrations (~11 – 15 wt%) and the western unit (Unit 2) shows comparatively lower FeO concentrations (~8 – 11 wt%) (Fig. 4a). The $TiO_2$ concentration for LL Unit 1 and Unit 2 is ~1 – 3.5 wt% and <1.5 wt%, respectively (Fig. 4b). We have utilized the CSFD technique to estimate the ages of these two units. The crater counting reveals that the basalts of the LL have been emplaced in at least two distinct episodes. The LL Unit 1 and LL Unit 2 give the model age of $3.3^{+0.06}_{-0.01}$ Ga for 350 m < D < 1.2 km and $2.1^{+0.2}_{-0.2}$ Ga for 250 m < D < 1 km (Fig. 5a, b). The older age of the LL Unit 1 is consistent with the results from Morota et al. (2011), which also suggests an age of ~3.3 Ga for the northern unit of the LL. The younger age of the LL Unit 2 suggests that this region has been volcanically active for a longer period than previously understood.

The FeO and $TiO_2$ concentration of the mare within the BB is ~9 – 15 wt% and <1.5 wt%, respectively. A small crater (diameter ~2 km) in the northern part of the BB crater floor has excavated the material having a slightly higher $TiO_2$ concentration of ~3.5 wt% (Fig. 4b). Crater counting yields an age of $3.4^{+0.05}_{-0.07}$ Ga (curve fitted for diameter ranges of 350 m < D < 1 km) for the mare emplaced within the BB crater (Fig. 5c), which is consistent with the age estimation from Morota et al. (2011). The estimated model age of the mare within the BB crater (~3.4 Ga) is equivalent to that of the mare emplaced within the LL Unit 1 (~3.3 Ga) within the uncertainty limits. The contemporaneous ages with similar FeO and $TiO_2$ concentrations suggest that both eruptions may have originated from the same magma source.

The LL Unit 1, LL Unit 2, and BB mare have a lower FeO concentration than the typical lunar mare basalts. Interestingly, these basalts of LL Unit 1, LL Unit 2, and BB also exhibit relatively higher $Al_2O_3$ concentrations of ~16 – 20 wt%, ~18 – 24 wt%, and ~20 – 23 wt%, respectively (Fig. 4c). According to the constraints provided by Kramer et al. (2008) on the FeO and $TiO_2$ concentrations, high-alumina (HA) basalts are characterized by relatively lower FeO (12 – 18 wt%) than typical mare basalts, very low to low $TiO_2$ content (1 – 5 wt%), and elevated $Al_2O_3$ (11 – 19 wt%). The basalts of the LL Unit 1 and BB fall within the FeO and $TiO_2$ limits suggested by Kramer et al. (2008), and also exhibit elevated $Al_2O_3$ concentration, which is consistent with the characteristics of HA basalts. The basalts of LL Unit 2 show exceptionally higher alumina content and reduced FeO concentrations. This could be due to the possible mixing with the surrounding highland anorthositic component, which may have further elevated the alumina content in the basalts of LL Unit 2. Additionally, studies based on analyses of HA basalts from Apollo 12, Apollo 14, and Luna 16 showed that HA basalt eruptions on the Moon span from ~4.2 to ~3.1 Ga (Cohen et al., 2001; Dasch et al., 1987; Nyquist et al., 1981; Papanastassiou & Wasserburg, 1971). Interestingly, our in-depth chronological and compositional analyses show that the HA basalts of the FS Basin have been emplaced during multiple volcanic episodes starting from ~3.4 Ga and extending till ~2.1 Ga, which indicates that the HA basalt eruptions on the Moon might have lasted for an extended period of time than previously reported. The small, fresh craters within both LL units and BB have excavated materials with relatively lower $Al_2O_3$ concentrations (~13 – 18 wt%), which still fall within the compositional range of HA basalts (Fig. 4c). A clear trade-off between $Al_2O_3$ and FeO is observed in the region where these small impacts have excavated lower $Al_2O_3$ materials, which have increased FeO concentrations by ~2 – 4 wt% compared to the surrounding basalts. The lower $Al_2O_3$/higher FeO materials excavated by small, fresh impacts reveal the presence of a compositionally distinct unit underlying the uppermost mare emplacements in LL and BB. The presence of different compositional layers indicates that episodic layered volcanism has occurred within the FS Basin.

For a detailed mineralogical characterization of the region, we have conducted an in-depth analysis of the reflectance spectra obtained from the $M^3$ data. The Band I and Band II center values for the LL Unit 1 lie between ~984-1079 nm and 1993-2269 nm, respectively. LL Unit 2 shows Band I and Band II center values ranging from ~1011-1043 nm and ~2061-2297 nm, respectively (Fig. 6a). These band center values are comparable to those of the HCP-bearing basalts. Most of the spectra collected from the LL Unit 1 and Unit 2 show a weak 1200 nm absorption band (See Appendix A Fig. A1 (c - f)), which suggests either a mixture of plagioclase and cpx, olivine and cpx, and/or very high calcium pyroxenes (Cheek et al., 2013; Klima et al., 2008). The presence weak 1200 nm absorption band is also indicative of a faster cooling of the basalts in the region (Klima et al., 2008). Spectra collected from the regions surrounding the small ~2.6 km diameter DHC located within the LL Unit 1 (Fig. 1a, b) show prominent signature of olivine (Fig. 7a, b, g). This further strengthens the possibility of olivine-mixing in the basalts of LL Unit 1 and LL Unit 2. The excavation depth of this DHC has been found to be ~218 m based on the relationship $d = 0.1D_{tc}$, where d is the excavation depth and the transient crater diameter $D_{tc} = 0.84D$ (for craters with Diameter (D) < 15 km) (Melosh, 1989; Pilkington & Hildebrand, 2003). This reveals the presence of a distinct olivine-rich compositional layer at a depth of ~218 m, which may extend into the subsurface of LL Unit 2. The source region of the identified olivine-rich mineralogy may indicate either localized magmatic intrusions within the subsurface or the incorporation of olivine-rich upper mantle material tapped during the formation of a large ~600 km diameter FS Basin. From Fig. 6b, it is observed that the spectral data points of LL Unit 1 and LL Unit 2 scatter away towards the right of the lunar olivine-cpx-opx (Ol-Cpx-Opx) mixing line. This suggests either an increasing Ti content in the basalts of this region or possible glass mixing with cpx (Zhang & Cloutis, 2021b). Additionally, the spectra obtained from the downrange of BB and the BB-LL transition zone show shallow Band I absorption and broader Band II absorption (Fig. 7c, d, g), which resembles the mature glassy soil mixed with pyroxenes from Apollo 11 (See Figure 6 from McCord et al. (1981)). The region showing the spectral signatures of glass mixing does not contain any clearly distinguishable vent structure that would support an explosive eruption resulting in the in-situ emplacement of pyroclastics and the formation of glassy materials. One possibility is that any vents associated with earlier explosive eruptions were subsequently buried beneath the later emplaced mare in LL, and no surface expression of vent structure is visible. This is consistent with the observation that most of the prominent signatures of glass mixed spectra are obtained from the area where mare emplacements are sparse, which allows the spectral detection of older glass-mixed material without getting completely covered by later mare emplacements. Another possibility is that the glassy component may not be volcanic, instead, it is impact melt that can be associated with the oblique BB impact. The study by Schultz et al. (1998) suggests that friction-generated melting can occur during a low-angle impact in the downrange direction of the crater formed by oblique impact, which aligns with the spatial distribution of glass signatures observed in this region. While the spectral signatures suggest the presence of glass-mixed materials in this region, confirming the origin of glassy material requires further detailed investigation.

The Band I and Band II center values for BB basalts lie between ~975-1090 nm and ~2042-2200 nm, respectively (Fig. 6a). The band center values of the basalts of BB are comparable to those of the LL Unit 1 and LL Unit 2, suggesting their similar HCP-rich compositions. Moreover, the spectra obtained from BB also show a weak, shallow 1200 nm absorption (See Appendix A Fig. A1 (a, b)), further suggesting a mixture of plagioclase and cpx, olivine and cpx, and/or very high-Ca pyroxenes (Cheek et al., 2013; Klima et al., 2008). The spectra obtained from the NNW crater-wall area of the BB show absorption at 1250 nm without any absorption at 1000 nm or 2000 nm (Fig. 7e, f, g). The 1250 nm absorption feature arises from electronic transitions in $Fe^{2+}$ that substitutes for $Ca^{2+}$ in irregular eight- or twelve-fold sites (Adams & Goullaud, 1978; Bell & Mao, 1973). This 1250 nm absorption is a characteristic spectral signature of a nearly pure plagioclase and corresponds to the 'purest anorthosite' (PAN) spectra (Ohtake et al., 2009). The detection of PAN in the BB crater wall is consistent with the previous global identifications of anorthositic exposures using SELENE, Diviner and $M^3$ data (e.g., Donaldson Hanna et al., 2012, 2014; Yamamoto et al., 2012). The identified locations of PAN-dominated lithology in BB lie along the proposed IDR of the FS Basin. Interestingly, similar PAN exposures have also been reported along the inner rings of other lunar basins such as Orientale, Neactaris, Grimaldi, and Humorum (Cheek et al., 2013; Donaldson Hanna et al., 2012; Donaldson Hanna et al., 2014; Ohtake et al., 2009; Srivastava & Gupta, 2013; Yamamoto et al., 2012), which further supports the existence of an inner depression ring in the FS Basin.

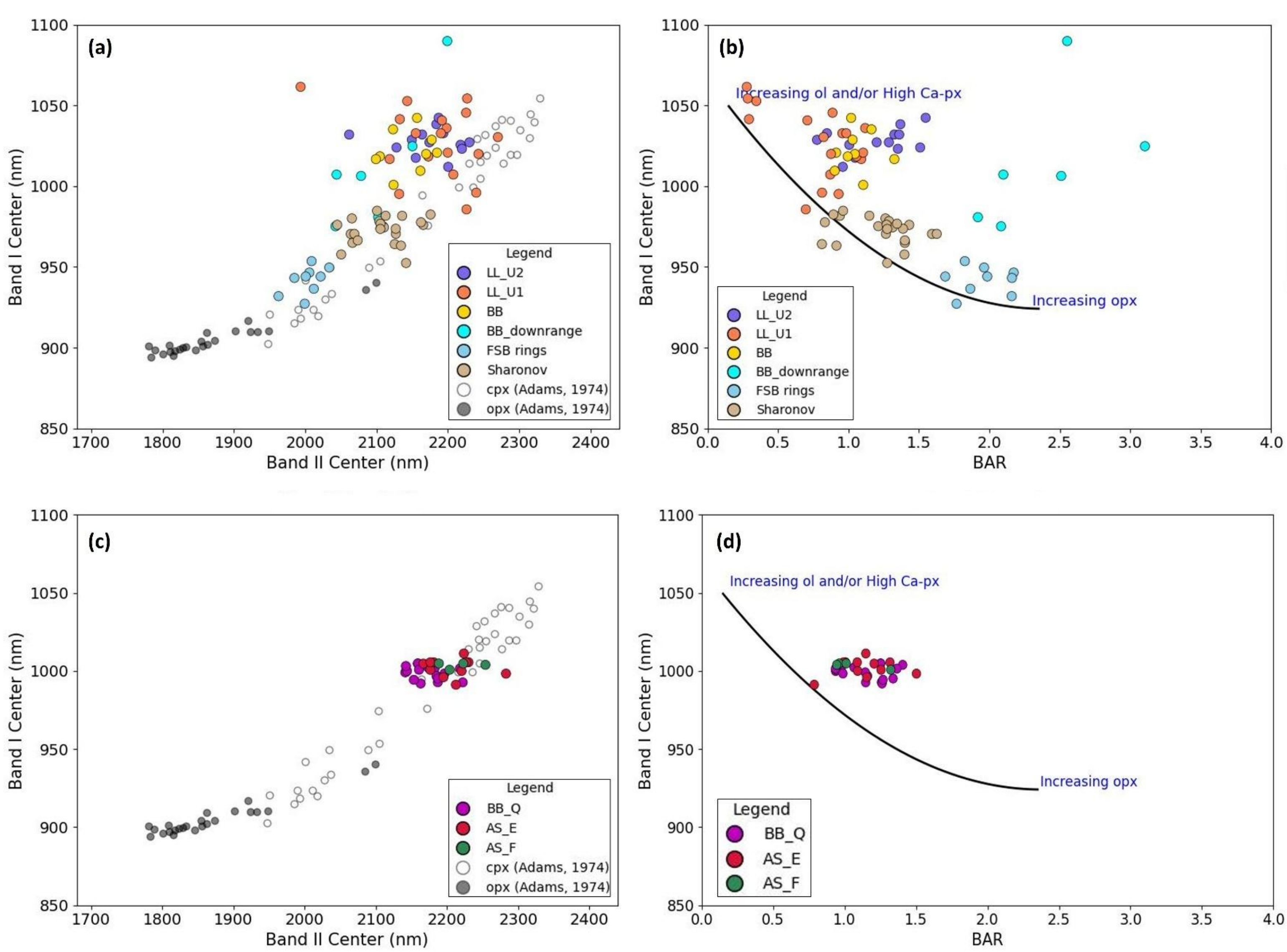

Fig. 6 – (a, c) BC I vs. BC II plots (modified after Adams (1974)) displaying the classification of pyroxenes into LCPs and HCPs, and (b, d) BC I vs. BAR plots showing the classification of minerals on the lunar Ol-Cpx-Opx mixing line adopted from *(Zhang & Cloutis, 2021a)*.

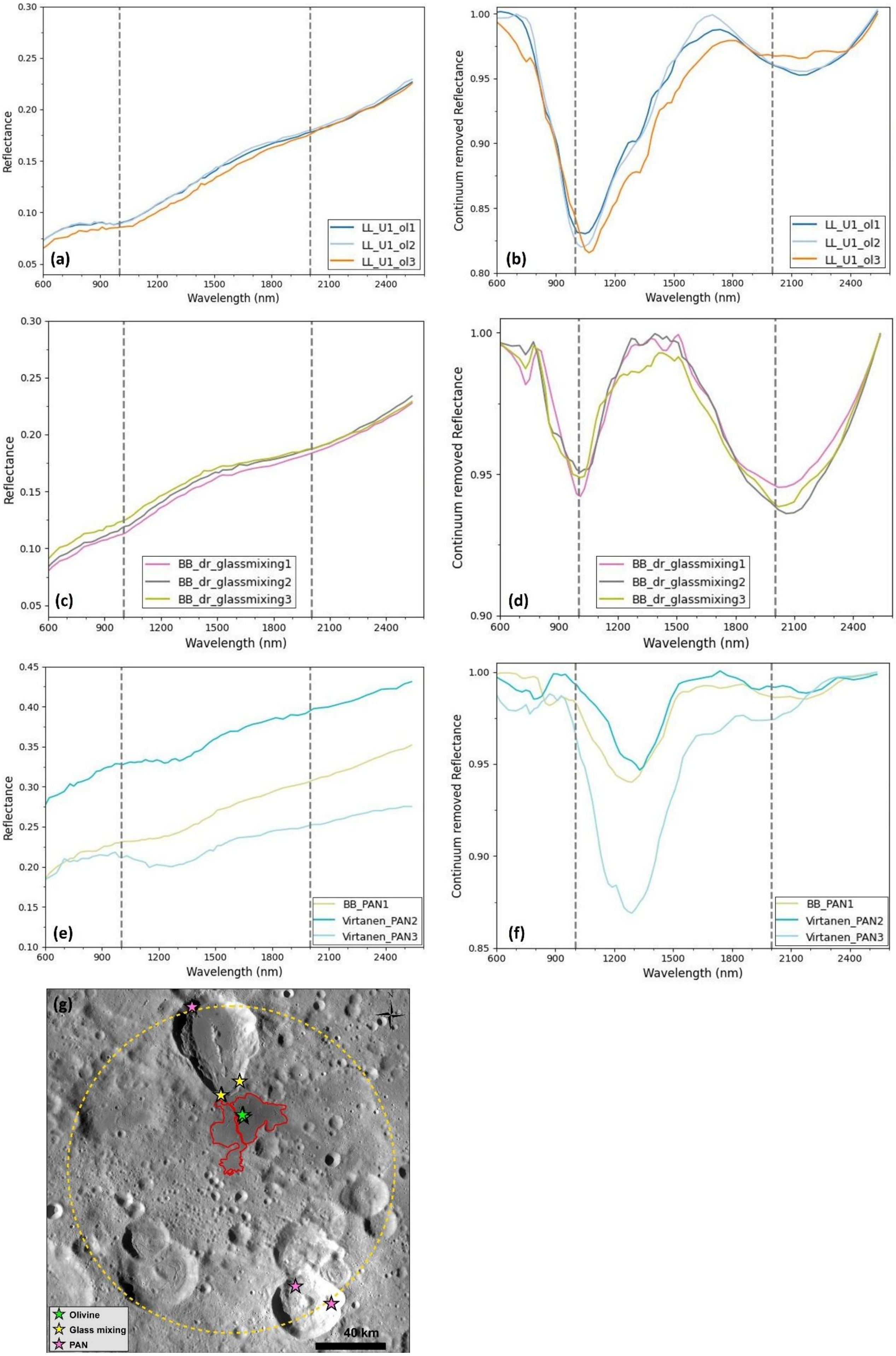

(a)
Reflectance
Wavelength (nm)
LL_U1_ol1
LL_U1_ol2
LL_U1_ol3
(b)
Continuum removed Reflectance
Wavelength (nm)
LL_U1_ol1
LL_U1_ol2
LL_U1_ol3
(c)
Reflectance
Wavelength (nm)
BB_dr_glassmixing1
BB_dr_glassmixing2
BB_dr_glassmixing3
(d)
Continuum removed Reflectance
Wavelength (nm)
BB_dr_glassmixing1
BB_dr_glassmixing2
BB_dr_glassmixing3
(e)
Reflectance
Wavelength (nm)
BB_PAN1
Virtanen_PAN2
Virtanen_PAN3
(f)
Continuum removed Reflectance
Wavelength (nm)
BB_PAN1
Virtanen_PAN2
Virtanen_PAN3
(g)
Olivine
Glass mixing
PAN
40 km

Fig. 7 – (a – f) Reflectance spectra and their corresponding continuum-removed spectra. (g) LRO WAC mosaic showing locations of the spectra plotted in (a – f), where green, yellow, and pink stars represent locations of olivine, glass-mixed materials, and PAN spectra, respectively. The dashed yellow circle marks the IDR of the FS Basin. Red polygons demarcate the boundaries of Lacus Luxuriae units.

### 4.2.2 Dark Mantle Deposits

The DMDs are primarily characterized by their very low albedo, smooth surfaces, and mantling relationship to the underlying surface features. The occurrence of DMDs is often attributed to the presence of explosively emplaced pyroclastic materials (Head, 1974). The craters Anderson E, Anderson F, and Buys-Ballot Q, situated in the central part of the FS Basin, have their floors partially covered by the DMDs, and show the development of the floor-fractured morphology (Fig. 1a, c, d). The Anderson E and Anderson F craters have previously been identified as potential DMDs (Besse et al., 2014; Gustafson et al., 2012) and linked to the presence of pyroclastics emplaced by explosive volcanic activity (Gustafson et al., 2012). In this study, we propose Buys-Ballot Q as a newly identified DMD (section 4.1) located in close proximity to the Andersons.

The DMDs within the craters Anderson E, Anderson F, and Buys-Ballot Q have FeO concentrations in the ranges of ~9 – 16 wt%, ~8.5 – 12.5 wt%, and ~9 – 13.5 wt%, respectively (Fig. 4a). The $TiO_2$ abundance of all three DMDs is <1.5 wt%, with exceptions at a few locations where small, fresh impacts have excavated the materials with elevated $TiO_2$ concentrations of ~3.5 – 4.8 wt% (Fig. 4b). The analysis of spectral parameters derived using the $M^3$ data shows that the BC I values of the DMDs Anderson E, Anderson F, and Buys-Ballot Q lie between ~991-1012 nm, ~1000-1013 nm, and 991-1008 nm, respectively. The BC II values of these DMDs range from ~2161-2283 nm, ~2188-2254 nm, and 2141-2223 nm,

respectively (Fig. 6c, d). The careful spectral analysis of the DMDs of Anderson E and Anderson F shows no signature of pyroclastic materials or glass, as indicated by the absence of reduced interband distance (Fig. 6c, Appendix A Fig. A2) and the lack of deviation from the lunar Ol-Cpx-Opx mixing line (Fig. 6d). Their mineralogy is consistent with the previously reported majorly HCP-rich basalts (Besse et al., 2014). Most of the locations from which Buys-Ballot Q spectra have been obtained exhibit a cpx-dominated mineral character. However, some locations exhibit very shallow Band II and concave inflection with weak absorption around 1200 nm (See Appendix A Fig. A2 (e, f)), indicative of either olivine, feldspathic materials, or glass mixing (Cheek et al., 2013; Klima et al., 2008).

#### 4.2.3 Non-mare regions

The non-mare regions within the FS Basin from where the spectra have been collected, include the rings of the FS Basin, and the craters Sharonov, Virtanen, and Virtanen F. The spectra obtained along the rings of the FS Basin have BC I and BC II values ranging from ~927-954 nm and ~1962-2034 nm, respectively, indicating the opx-rich exposures along the basin rings (Fig. 6a, Appendix A Fig. A3). The Band I and Band II center values for Sharonov crater lie between ~952-985 nm and ~2044-2176 nm, respectively (Fig. 6a). The band center values of the Sharonov crater are comparable and typical of the LCP-bearing cpx-rich lithology. It is clear from Fig. 6b that the lithology of the Sharonov crater has a higher Ca-content than the materials exposed along the FS Basin rings. The spectra obtained from the Virtanen and Virtanen F craters show prominent absorption at 1250 nm without any absorption at 1000 nm and 2000 nm (Fig. 7e, f), indicating the PAN exposures. Notably, the regions that have exposed PAN lithology in and around Virtanen crater also spatially align with the proposed inner depression ring of the FS Basin (Fig. 7g). The detected PAN exposure from the Virtanen crater is consistent with the previously reported PAN-bearing lithologies by Yamamoto et al. (2012).

The occurrence of opx and PAN-bearing lithologies along the rings of the FS Basin suggests that the basin formation process has sampled materials from the lower-crustal depths.

The Virtanen crater has excavated Fe-depleted material (<2.5 wt% FeO) from the subsurface (Fig. 4a), implying that at the time of Virtanen crater formation, the magma had not yet intruded the subsurface; otherwise, the crater would have excavated Fe-enriched material. To determine how long the subsurface remained devoid of magmatic activity after the formation of the FS Basin, we performed crater counting on the ejecta of the Virtanen crater and estimated its age. The ejecta blanket of the Virtanen crater shows multiple resurfacing events due to the mixing by later-formed crater ejecta. Here, our objective is to constrain the timeline of the initiation of the magmatic activity in the FS Basin, we are only concerned about the oldest exposure of the Fe-depleted subsurface material within the FS Basin. The oldest age for the Fe-poor Virtanen ejecta is estimated to be $3.8^{+0.04}_{-0.05}$ Ga (curve fitted for diameter (D) ranges of 1.1 km $< D <$ 3 km) (Fig. 5d), which provides a minimum upper bound on the timeline before which the subsurface of the FS Basin was not modified by any magmatic intrusions. Thus, magma intruded the subsurface at least after ~350 Ma of the formation of the FS Basin between ~3.8 Ga to ~3.4 Ga.

**4.3 Controls on the volcanism in the Freundlich-Sharonov Basin**

The FS Basin exhibits a spatially restricted and volumetrically limited volcanic activity confined within the central depression and along the proposed inner depression ring of the basin. The morphological investigation has shown that in the central region, the magma has stalled within the subsurface, forming localized dikes and sills. The chronological analysis has shown at least two episodes of volcanism in the region, with the youngest eruption occurring at ~2.1 Ga. It is essential to comprehend the factors governing volcanism in the FS Basin, which has formed on a thicker farside crust, away from heat-producing elements.

Crustal thickness has been suggested as a dominant factor controlling the volcanic eruptions on the Moon. In the regions with thicker crust, magma cannot easily rise to the surface and instead forms intrusive magmatic bodies (Head & Wilson, 2017). The central parts of the FS Basin exhibit crustal thickness of ~10 km (Wieczorek et al., 2013), implying significant crustal thinning relative to the pre-impact farside crust due to the basin-forming impact. Although FS Basin has a lower crustal thickness, comparable to several nearside basins (e.g., Orientale, Fecunditatis, Imbrium, Humorum), the mare emplacements within the FS Basin are sparse. Moreover, the mare emplacements are absent even in the regions along the other two rings of the FS Basin and at intersections with the adjacent FJ Basin rings, despite the likely presence of extensive fault systems capable of facilitating magma ascent (Nahm et al., 2013; Panwar et al., 2025).

The majority of mare volcanism on the Moon is observed to have confined within or closely associated with impact basins. These observations have suggested a possible link between the formation of impact basins and the onset of volcanism (Elkins-Tanton et al., 2004; Ghods & Arkani-Hamed, 2007; Panwar et al., 2023; Whitten et al., 2011). In addition to impact-modified crustal morphologies, studies have proposed impact-induced decompression melting and localized mantle convections as possible mechanisms producing magma (Elkins-Tanton & Hager, 2005; Ghods & Arkani-Hamed, 2007). The FS Basin is dated as a pN/N impact basin formed ~4.1 Ga ago. However, crater-counting results from this study suggest that mafic magma had not intruded until at least ~3.8 Ga. This lag between the basin formation age and the magma intrusion age indicates that the decompression melting may not be responsible for magma generation, but does not rule out the possibility of impact-induced localized mantle convection as a possible mechanism that might have aided in magma production following the FS Basin-forming impact.

All surface eruptions within the FS Basin have a characteristic composition of HA basalts with very low to low Ti contents. The dominance of HA basalts may suggest a relationship to the thick pre-impact anorthositic crustal column. Previous studies have proposed that basins formed in thicker crusts preferentially host low-Ti, high-Al magmas, as the crustal thickness may influence the composition of erupted magma (Kramer et al., 2008; Wilhelms et al., 1987). Magmas with these high-Al and low-Ti compositions have very low densities (~2800 – 2850 kg/m$^3$; Wieczorek et al., 2001), allowing them to achieve positive buoyancy compared to the anorthositic crust and erupt onto the surface. In contrast, denser magma with a different composition would stall within the subsurface at a shallow neutral buoyancy horizon (Lister, 1990), consistent with the morphologic and morphometric evidence for subsurface magmatic intrusions in the central parts of the FS Basin. Although theoretical and experimental modeling have shown that under a given magma production rate, stalled dikes may progressively overshoot the neutral buoyancy horizon as they grow vertically (Lister & Kerr, 1991; Wieczorek et al., 2001), in the FS Basin, we do not see extensive surface eruptions, and the majority of magma appears to have stalled within the subsurface. A higher intrusive to extrusive ratio for farside magmatism is also observed by previous studies such as Broquet & Andrews-Hanna (2024), and Laneuville et al. (2013).

The limited extent of surface eruption observed in the FS Basin can therefore be explained by three possibilities: (a) The ascent and eruption of magma was controlled by the composition and, in turn, the buoyancy of the magma. Only the high-Al magma with very low density could rise to the surface, whereas the denser magma with a different composition could not achieve positive buoyancy and did not extrude. The excavation of olivine-rich material by a DHC in LL Unit 1 and higher FeO materials exposed by small impacts in LL and BB support the possible compositional heterogeneity in the magma sources. (b) The magma production in this region is insufficient to produce enough volume of magma that can aid the vertical growth of the dike stalled at shallow crustal levels, which can eventually overshoot the neutral buoyancy horizon and erupt onto the surface (Taisne et al., 2011; Taisne & Tait, 2009; Wilson & Head, 2017). Based on the volumes of extruded mare basalts, several studies, including Morota et al. (2009), Taguchi et al. (2017), and Wieczorek et al. (2001), have suggested that the quantity of magma produced within the farside mantle is ~3 – 20 times less than that on the nearside. However, these estimates are based primarily on the erupted mare volumes and do not account for stalled subsurface intrusions. Therefore, it remains uncertain whether the magma production in the farside mantle is indeed low, and if it is the case, then by what magnitude. (c) Vertically propagating dikes were not able to maintain sufficient magmatic pressure within the column to penetrate the surface. The proposed IDR of the FS Basin is superimposed by numerous impact craters, each generating a low-density brecciated lens below their crater floors. The deep-reaching faults associated with IDR can help overpressurized magma columns ascend through the crust, forming ring dikes and producing a pressure gradient. However, when the magma rising through such dikes encounters the brecciated zones of small-scale fractures beneath the superposed craters, the magmatic driving pressure dissipates, and it is not able to continue the vertical ascent. As the pressure in such stalled dikes exceeds the lithostatic pressure, the dikes begin to propagate laterally, forming a sill or laccolith (Thorey & Michaut, 2014). The formation of a sill or laccolith beneath the crater can result in upbowed or fractured crater floors as observed in Virtanen, Buys-Ballot Q, Anderson E, and Anderson F craters. Additionally, the detailed mineralogical analysis carried out in this study indicates that the vents formed within these craters due to floor-fracturing are not associated with pyroclastic material, representing the passive degassing of the intrusion from the deflation of under-pressurized magmatic foam (Head & Wilson, 2017; Jozwiak et al., 2015). These inferences suggest that the magma column may not have been pressurized sufficiently to sustain vertical propagation to the surface and instead stalled at shallow crustal levels.

## 5. Conclusions

This study investigates the geological evolution and magmatism of the pN/N-aged Freundlich-Sharonov Basin (~600 km diameter), located on the farside of the Moon. A variety of high-resolution remote sensing data from missions such as Chandrayaan-1, Kaguya, and LRO have been used to conduct a detailed morphological, compositional, and chronological analysis of this region, which helped constrain the controls on the volcanism in the FS Basin.

Morphometric analysis suggests the presence of subsurface magmatic intrusions in the central region of the FS Basin. The floor-fractured morphology observed in the craters Anderson E, Anderson F, Buys-Ballot Q, and the uplifted crater floor of the Virtanen are attributed to the presence of shallow magmatic intrusions and sill formation in the subsurface. We propose the presence of a previously unrecognized Inner Depression Ring (IDR) with a diameter of ~192 km, supported by the identified spatially aligned subsurface magmatic intrusions as well as PAN and opx exposures.

A spectral investigation conducted using hyperspectral data from $M^3$ has revealed the presence of diverse mineralogy across the FS Basin, comprising orthopyroxenes, clinopyroxenes, olivine, PAN, and possible glass-mixed compositions. The identified DMDs associated with Anderson E, Anderson F, and Buys-Ballot Q craters exhibit compositions indicative of HCP-rich basalts rather than pyroclastics, which is consistent with the results from Besse et al. (2014). Based on the variation in Kaguya-derived FeO concentrations, the mare emplacements within Lacus Luxuriae (LL) have been classified into two distinct units, with Unit 1 having higher FeO concentrations with respect to the Unit 2. The mare emplaced within Buys-Ballot (BB), and both LL units show the mineralogy dominated by high-Ca clinopyroxenes. Spectra from the BB-LL transition zone and BB downrange direction suggest the possibility of mixing of pyroxenes with glassy components. It remains unclear whether this glass is formed as a result of explosive volcanism or due to impact melting. A ~2.6 km diameter crater within LL Unit 1 has excavated olivine-rich materials from a depth of ~218 m, revealing compositional layering in the region. The olivine-rich materials may originate from localized magmatic intrusions or represent the upper mantle material excavated during the basin-forming impact, which was later overlain by subsequent volcanic episodes. Exposures of PAN and orthopyroxenes along the basin rings indicate excavation of the lower crustal lithologies during the basin formation.

Chronological analysis shows that the FS Basin has experienced at least two distinct episodes of volcanism at ~3.4 – 3.3 Ga in the BB and LL Unit 1, and at ~2.1 Ga within the LL Unit 2. The emplacement of late-stage basalts in LL Unit 2, along with the preserved boulder tracks and lobate scarps, indicates that the region has remained tectonically and volcanically active until very recent geological times.

The mare exposures within the central FS Basin are predominantly composed of exceptionally high-alumina (HA) basalts (~16 – 24 wt% $Al_2O_3$) with very low to low $TiO_2$ content, which can be an important in-situ resource utilization (ISRU) feedstock for aluminium and oxygen extraction. In the LL Unit 2, the HA basalts have emplaced ~2.1 Ga, indicating that HA basalt eruptions might have continued for a longer geological time period than previously understood. The dominance of HA basalts may suggest a relationship to the thick pre-impact anorthositic crustal column, which could have influenced the composition of the erupted magma.

We have suggested three possibilities explaining why the intruded magma might not be able to erupt onto the surface: (i) Magma with denser compositions could not achieve positive buoyancy, which restricted its ascent to the surface. (ii) Reduced magma production in the farside mantle limited the ability of stalled dikes to grow vertically and overshoot the neutral buoyancy horizon. (iii) Numerous superposed impacts along the IDR produced low-density brecciated zones beneath those craters, preventing the magma column from maintaining sufficient pressure to sustain vertical propagation and penetrate the surface. Thus, the study enhances our understanding of the volcano-tectonic evolution of the FS Basin and highlights the importance of magma properties in conjunction with impact-modified crustal structures, which influence the sites of magma extrusion on the lunar surface.

## Acknowledgements

Tvisha Kapadia duly acknowledges the financial support provided by the Department of Science and Technology (DST), Government of India, under the INSPIRE Fellowship Scheme. The authors N.P., N.S., M.B., R.S., and A.B. are funded by the Department of Space, Government of India. We thank the Chandrayaan-1 $M^3$, LRO, and Kaguya MI teams for providing calibrated data through publicly accessible portals.

## Appendix A: $M^3$ Reflectance Spectra of the FS Basin

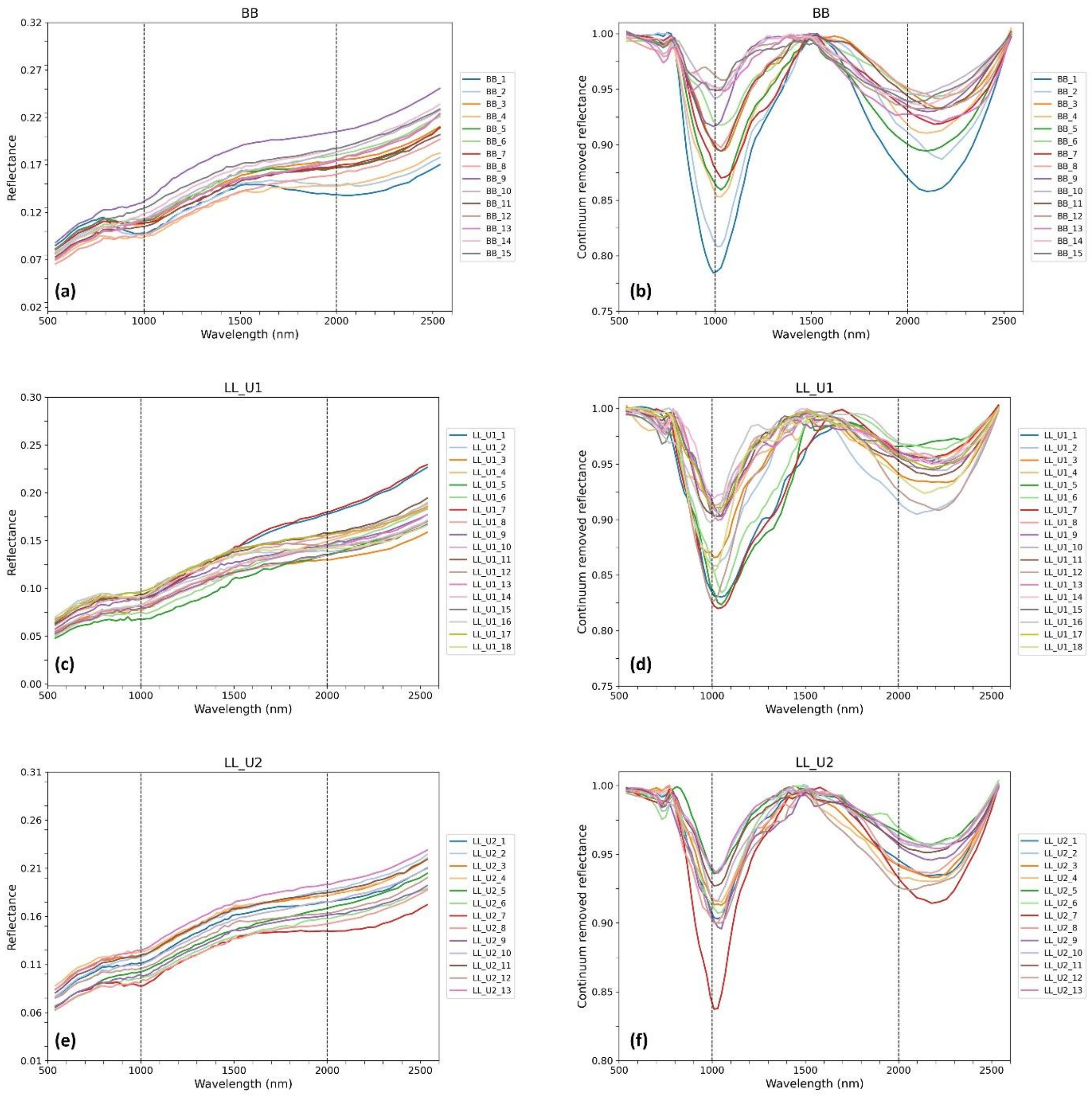

Fig. A1 - Reflectance spectra of mare units within FS Basin and their corresponding continuum-removed profiles. BB: Buys-Ballot, LL U1: Lacus Luxuriae Unit 1, LL U2: Lacus Luxuriae Unit 2.

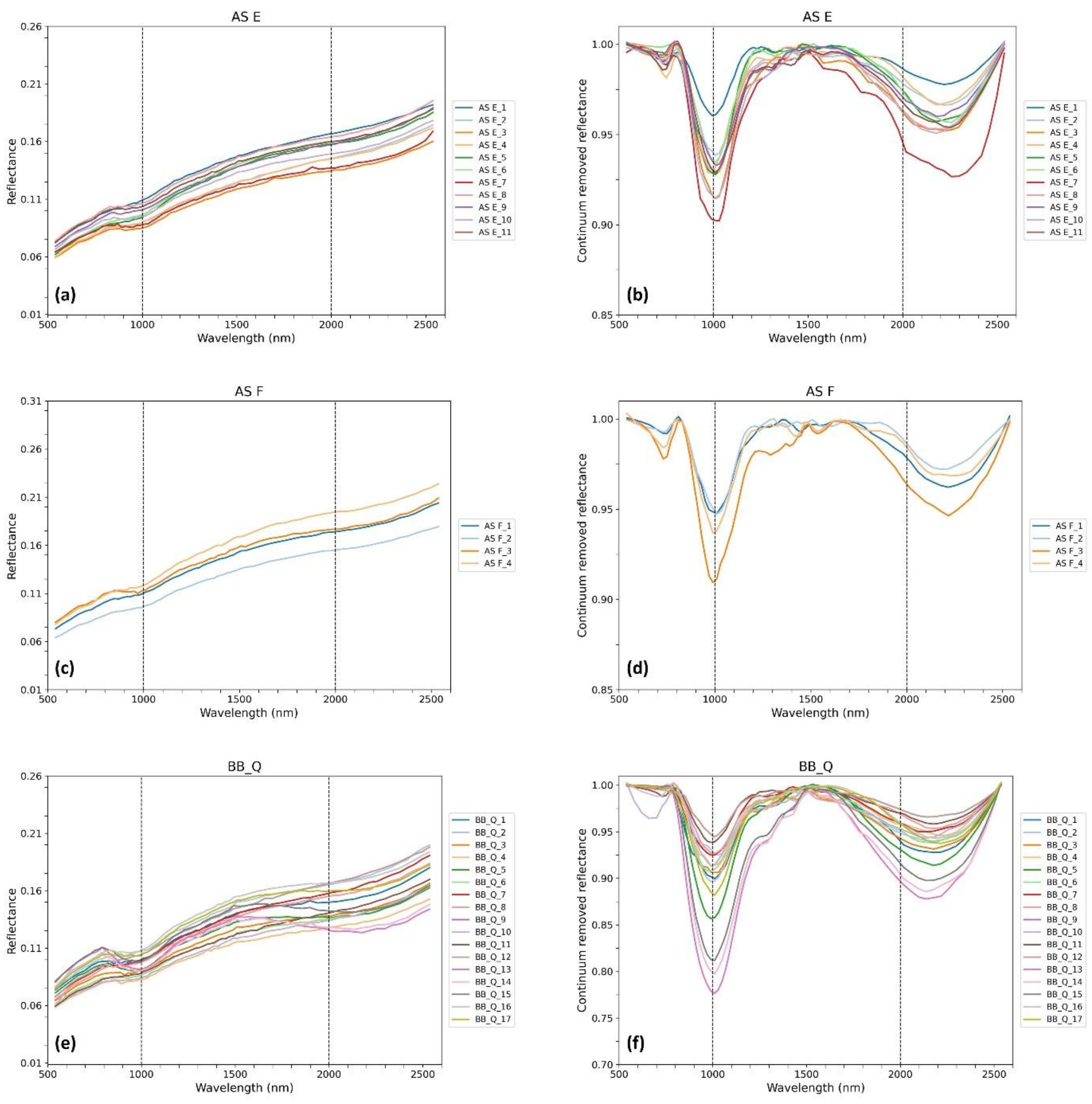

Fig. A2 - Reflectance spectra of DMDs within FS Basin and their corresponding continuum-removed profiles. AS E: Anderson E, AS F: Anderson F, BB Q: Buys-Ballot Q.

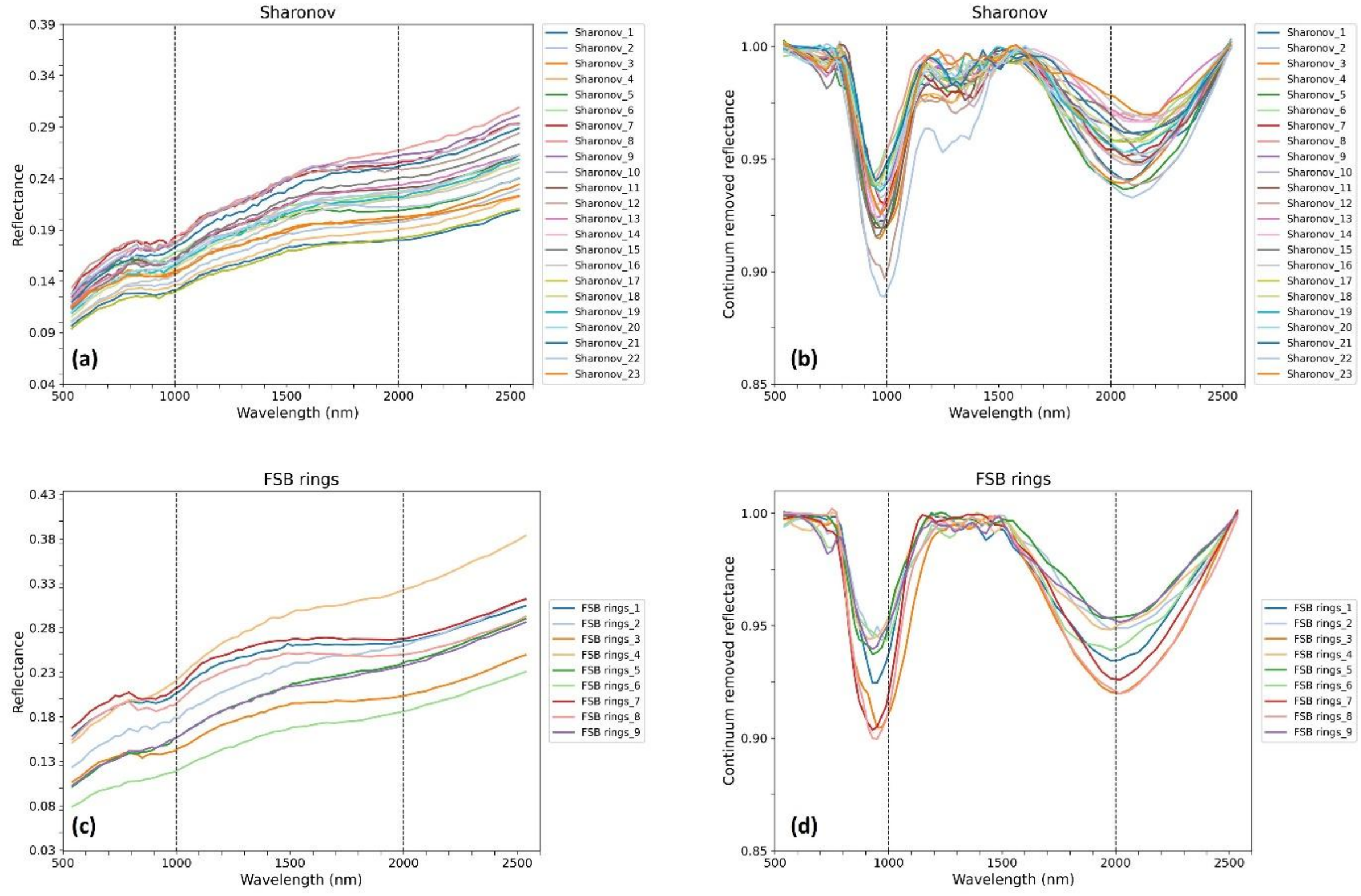

Fig. A3 - Reflectance spectra of non-mare regions within FS Basin and their corresponding continuum-removed profiles.

## Appendix B: Locations and Band Parameters of the Reflectance Spectra Used in the Study

Table B1 - Locations and band parameters of the spectral points

| Spectra name | Latitude | Longitude | BC I | BC II | BAR |
|---|---|---|---|---|---|
| BB_1 | 21.479 | 174.832 | 1000.75 | 2123.87 | 1.105 |
| BB_2 | 21.273 | 174.928 | 1009.51 | 2161.4 | 0.902 |
| BB_3 | 21.029 | 175.014 | 1028.71 | 2177.24 | 1.029 |
| BB_4 | 21.427 | 174.511 | 1018.39 | 2104.67 | 0.995 |
| BB_5 | 21.690 | 174.813 | 1016.71 | 2099.07 | 1.326 |
| BB_6 | 21.695 | 174.636 | 1035.15 | 2123.07 | 1.163 |
| BB_7 | 20.645 | 174.717 | 1042.23 | 2156.68 | 1.019 |
| BB_8 | 21.192 | 174.636 | 1020.71 | 2184.68 | 0.912 |
| BB_9 | 20.070 | 175.024 | 980.78 | 2102.27 | 1.919 |
| BB_10 | 19.868 | 174.966 | 1007.15 | 2043.95 | 2.099 |
| BB_11 | 20.003 | 174.818 | 1019.91 | 2169.4 | 1.045 |
| BB_12 | 20.017 | 175.278 | 1024.71 | 2150.28 | 3.104 |
| BB_13 | 20.012 | 175.278 | 1089.79 | 2199.08 | 2.552 |
| BB_14 | 19.859 | 174.947 | 1006.31 | 2078.27 | 2.511 |
| BB_15 | 20.137 | 175.321 | 975.15 | 2042.42 | 2.084 |
| LL_U1_1 | 19.470 | 175.412 | 1052.67 | 2142.92 | 0.346 |
| LL_U1_2 | 19.796 | 175.772 | 1016.75 | 2118.27 | 1.090 |
| LL_U1_3 | 19.490 | 175.762 | 995.95 | 2239.64 | 0.812 |
| LL_U1_4 | 19.408 | 175.556 | 1040.71 | 2191.64 | 0.708 |
| LL_U1_5 | 19.413 | 175.456 | 1061.47 | 1993.54 | 0.278 |
| LL_U1_6 | 19.413 | 175.422 | 1054.27 | 2226.92 | 0.285 |
| LL_U1_7 | 19.461 | 175.398 | 1041.43 | 2132.52 | 0.294 |
| LL_U1_8 | 19.288 | 175.350 | 1019.91 | 2242.77 | 0.876 |
| LL_U1_9 | 19.178 | 175.441 | 1045.43 | 2225.48 | 0.887 |
| LL_U1_10 | 18.938 | 175.441 | 1020.71 | 2199.72 | 1.104 |
| LL_U1_11 | 19.542 | 175.264 | 1018.35 | 2172.52 | 1.056 |
| LL_U1_12 | 19.648 | 175.168 | 1007.19 | 2207.56 | 0.871 |
| LL_U1_13 | 19.576 | 175.523 | 1030.35 | 2269.97 | 0.823 |
| LL_U1_14 | 19.312 | 175.758 | 1035.95 | 2198.04 | 1.119 |
| LL_U1_15 | 19.418 | 175.801 | 1032.71 | 2154.92 | 0.957 |
| LL_U1_16 | 19.523 | 176.136 | 985.62 | 2226.04 | 0.697 |
| LL_U1_17 | 19.010 | 175.484 | 1032.71 | 2190.12 | 0.984 |
| LL_U1_18 | 19.619 | 176.208 | 995.15 | 2131.63 | 0.930 |
| | | | | | |
| LL_U2_1 | 18.401 | 175.077 | 1027.11 | 2230.13 | 1.200 |
| LL_U2_2 | 18.415 | 175.048 | 1023.91 | 2127.87 | 1.509 |
| LL_U2_3 | 18.195 | 175.115 | 1038.31 | 2183.72 | 1.368 |
| LL_U2_4 | 18.152 | 175.105 | 1042.31 | 2187 | 1.548 |
| LL_U2_5 | 18.924 | 174.942 | 1031.95 | 2163.8 | 1.326 |
| LL_U2_6 | 19.183 | 174.995 | 1028.67 | 2149.4 | 0.777 |
| LL_U2_7 | 19.298 | 175.269 | 1011.95 | 2200.36 | 0.960 |
| LL_U2_8 | 19.657 | 175.129 | 1017.51 | 2154.92 | 1.047 |
| LL_U2_9 | 19.173 | 175.192 | 1032.71 | 2193.48 | 0.847 |
| LL_U2_10 | 18.737 | 175.302 | 1025.59 | 2217.64 | 1.006 |
| LL_U2_11 | 18.521 | 175.120 | 1027.11 | 2174.04 | 1.287 |
| LL_U2_12 | 18.468 | 175.264 | 1031.91 | 2061.55 | 1.360 |
| LL_U2_13 | 18.368 | 175.403 | 1023.11 | 2219.64 | 1.351 |
| | | | | | |
| AS E_1 | 17.332 | 173.739 | 1000.02 | 2220.44 | 1.092 |
| AS E_2 | 17.226 | 173.633 | 1005.59 | 2230.04 | 1.006 |
| AS E_3 | 17.222 | 173.614 | 1004.71 | 2166.83 | 1.207 |
| AS E_4 | 17.222 | 173.590 | 991.22 | 2212.44 | 0.788 |
| AS E_5 | 17.087 | 173.715 | 995.99 | 2194.92 | 1.155 |
| AS E_6 | 17.073 | 173.739 | 1005.51 | 2226.84 | 1.318 |
| AS E_7 | 17.006 | 173.653 | 998.31 | 2282.77 | 1.503 |
| AS E_8 | 16.891 | 173.633 | 1000.75 | 2176.44 | 1.255 |
| AS E_9 | 17.193 | 173.456 | 1005.55 | 2181.32 | 1.089 |
| AS E_10 | 17.183 | 173.470 | 1005.55 | 2176.6 | 0.993 |
| AS E_11 | 16.953 | 173.312 | 1011.19 | 2223.64 | 1.149 |
| | | | | | |
| AS F_1 | 16.560 | 174.108 | 1000.79 | 2203.56 | 1.323 |

| AS F_2 | 16.536 | 174.099 | 1004.79 | 2188.52 | 0.962 |
|---|---|---|---|---|---|
| AS F_3 | 16.047 | 173.686 | 1004.82 | 2222.77 | 1.012 |
| AS F_4 | 16.056 | 173.720 | 1003.95 | 2253.89 | 0.945 |
| | | | | | |
| BB_Q_1 | 19.461 | 173.432 | 992.79 | 2187 | 1.258 |
| BB_Q_2 | 19.418 | 173.283 | 1003.91 | 2178.04 | 1.407 |
| BB_Q_3 | 19.403 | 173.288 | 1001.55 | 2182.28 | 1.366 |
| BB_Q_4 | 19.408 | 173.298 | 992.71 | 2222.04 | 1.148 |
| BB_Q_5 | 19.643 | 173.010 | 996.79 | 2185.24 | 1.161 |
| BB_Q_6 | 19.274 | 172.804 | 991.99 | 2163 | 1.263 |
| BB_Q_7 | 19.279 | 172.799 | 994.35 | 2153.4 | 1.269 |
| BB_Q_8 | 19.418 | 172.780 | 1002.31 | 2173.24 | 1.067 |
| BB_Q_9 | 19.509 | 172.708 | 1004.79 | 2159 | 1.252 |
| BB_Q_10 | 19.677 | 172.804 | 999.11 | 2141.4 | 1.145 |
| BB_Q_11 | 19.677 | 172.799 | 995.19 | 2187 | 1.339 |
| BB_Q_12 | 19.509 | 172.703 | 1004.79 | 2159 | 1.252 |
| BB_Q_13 | 19.595 | 172.617 | 999.95 | 2143.72 | 0.937 |
| BB_Q_14 | 19.595 | 172.622 | 1003.19 | 2142.2 | 0.981 |
| BB_Q_15 | 19.600 | 172.622 | 1000.79 | 2160.6 | 0.943 |
| BB_Q_16 | 19.600 | 172.636 | 998.39 | 2196.44 | 0.989 |
| BB_Q_17 | 19.168 | 172.507 | 1001.59 | 2218.04 | 0.937 |
| | | | | | |
| Sharonov_1 | 12.920 | 173.576 | 966.42 | 2073.55 | 1.400 |
| Sharonov_2 | 12.863 | 173.576 | 976.75 | 2105.4 | 1.345 |
| Sharonov_3 | 12.681 | 173.653 | 976.06 | 2044.82 | 1.271 |
| Sharonov_4 | 12.537 | 173.748 | 973.5 | 2105.31 | 1.275 |
| Sharonov_5 | 12.196 | 173.418 | 970.42 | 2063.79 | 1.593 |
| Sharonov_6 | 12.220 | 173.346 | 957.7 | 2050.42 | 1.397 |
| Sharonov_7 | 12.225 | 172.703 | 980.02 | 2065.39 | 1.262 |
| Sharonov_8 | 12.235 | 172.713 | 952.42 | 2141.32 | 1.275 |
| Sharonov_9 | 11.990 | 172.334 | 976.02 | 2165.32 | 1.431 |
| Sharonov_10 | 12.019 | 172.449 | 978.35 | 2103.87 | 1.288 |
| Sharonov_11 | 11.985 | 172.497 | 981.66 | 2112.67 | 1.148 |
| Sharonov_12 | 11.942 | 172.511 | 975.22 | 2108.67 | 1.211 |
| Sharonov_13 | 12.968 | 173.859 | 964.1 | 2125.4 | 0.811 |
| Sharonov_14 | 12.896 | 173.902 | 981.62 | 2135.8 | 0.938 |
| Sharonov_15 | 12.872 | 173.916 | 963.18 | 2134.2 | 0.915 |
| Sharonov_16 | 12.786 | 173.868 | 982.42 | 2175.64 | 0.893 |
| Sharonov_17 | 11.894 | 173.720 | 964.9 | 2066.27 | 1.400 |
| Sharonov_18 | 11.784 | 173.053 | 970.5 | 2127 | 1.266 |
| Sharonov_19 | 11.784 | 172.885 | 970.46 | 2068.83 | 1.626 |
| Sharonov_20 | 11.372 | 172.785 | 973.62 | 2126.83 | 1.387 |
| Sharonov_21 | 11.818 | 172.588 | 974.46 | 2110.27 | 1.314 |
| Sharonov_22 | 11.904 | 172.401 | 984.75 | 2100.6 | 0.962 |
| Sharonov_23 | 11.070 | 173.912 | 977.62 | 2162.2 | 0.833 |
| | | | | | |
| FSB rings_1 | 24.030 | 175.053 | 946.5 | 2005.79 | 2.173 |
| FSB rings_2 | 21.877 | 184.796 | 949.62 | 2033.62 | 1.963 |
| FSB rings_3 | 7.128 | 174.099 | 953.58 | 2008.9 | 1.826 |
| FSB rings_4 | 12.091 | 168.076 | 931.9 | 1962.58 | 2.159 |
| FSB rings_5 | 13.913 | 164.691 | 927.22 | 1999.3 | 1.767 |
| FSB rings_6 | 25.728 | 168.508 | 943.18 | 1984.82 | 2.160 |
| FSB rings_7 | 24.035 | 175.048 | 936.46 | 2012.02 | 1.865 |
| FSB rings_8 | 17.399 | 180.380 | 944.06 | 2021.62 | 1.985 |
| FSB rings_9 | 12.484 | 180.840 | 944.06 | 2000.66 | 1.687 |

**References**

Adams, J. B., & Goullaud, L. H. (1978). Plagioclase feldspars: Visible and near infrared diffuse reflectance spectra as applied to remote sensing. *Lunar and Planetary Science Conference Proceedings*, *3*, 2901–2909.

Adams, J. B. (1974). Visible and near-infrared diffuse reflectance spectra of pyroxenes as applied to remote sensing of solid objects in the solar system. *Journal of Geophysical Research (1896-1977)*, *79*(32), 4829–4836. https://doi.org/https://doi.org/10.1029/JB079i032p04829

Andrews-Hanna, J. C., Head, J. W., Johnson, B. C., Keane, J. T., Kiefer, W. S., McGovern, P. J., Neumann, G. A., Wieczorek, M. A., & Zuber, M. T. (2018). Ring faults and ring dikes around the Orientale basin on the Moon. *Icarus*, *310*, 1–20. https://doi.org/https://doi.org/10.1016/j.icarus.2017.12.012

Baker, D. M. H., & Head, J. W. (2013). New morphometric measurements of craters and basins on Mercury and the Moon from MESSENGER and LRO altimetry and image data: An observational framework for evaluating models of peak-ring basin formation. Planetary and Space Science, https://doi.org/https://doi.org/10.1016/j.pss.2013.07.003

Barker, M. K., Mazarico, E., Neumann, G. A., Zuber, M. T., Haruyama, J., & Smith, D. E. (2016). A new lunar digital elevation model from the Lunar Orbiter Laser Altimeter and SELENE Terrain Camera. *Icarus*, *273*, 346–355. https://doi.org/https://doi.org/10.1016/j.icarus.2015.07.039

Bell, P. M., & Mao, H. K. (1973). Optical and chemical analysis of iron in Luna 20 plagioclase. *Geochimica et Cosmochimica Acta*, *37*(4), 755–759. https://doi.org/https://doi.org/10.1016/0016-7037(73)90172-5

Besse, S., Sunshine, J. M., & Gaddis, L. R. (2014). Volcanic glass signatures in spectroscopic survey of newly proposed lunar pyroclastic deposits. *Journal of Geophysical Research: Planets*, *119*(2), 355–372. https://doi.org/https://doi.org/10.1002/2013JE004537

Besse, S., Sunshine, J., Staid, M., Boardman, J., Pieters, C., Guasqui, P., Malaret, E., McLaughlin, S., Yokota, Y., & Li, J.-Y. (2013). A visible and near-infrared photometric correction for Moon Mineralogy Mapper (M3). *Icarus*, *222*(1), 229–242. https://doi.org/https://doi.org/10.1016/j.icarus.2012.10.036

Boardman, J. W., Pieters, C. M., Green, R. O., Lundeen, S. R., Varanasi, P., Nettles, J., Petro, N., Isaacson, P., Besse, S., & Taylor, L. A. (2011). Measuring moonlight: An overview of the spatial properties, lunar coverage, selenolocation, and related Level 1B products of the Moon Mineralogy Mapper. *Journal of Geophysical Research: Planets*, *116*(E6). https://doi.org/https://doi.org/10.1029/2010JE003730

Broquet, A., & Andrews-Hanna, J. C. (2024). A volcanic inventory of the Moon. *Icarus*, *411*, 115954. https://doi.org/https://doi.org/10.1016/j.icarus.2024.115954

Burns, R. G. (1981). Intervalence Transitions in Mixed Valence Minerals of Iron and Titanium. *Annual Review of Earth and Planetary Sciences*, *9*(Volume 9, 1981), 345–383. https://doi.org/https://doi.org/10.1146/annurev.ea.09.050181.002021

Burns, R. G. (1993). Mineralogical Applications of Crystal Field Theory. In *Cambridge Topics in Mineral Physics and Chemistry* (2nd ed.). Cambridge University Press. https://doi.org/DOI: 10.1017/CBO9780511524899

Cheek, L. C., Donaldson Hanna, K. L., Pieters, C. M., Head, J. W., & Whitten, J. L. (2013). The distribution and purity of anorthosite across the Orientale basin: New perspectives from Moon Mineralogy Mapper data. *Journal of Geophysical Research: Planets*, *118*(9), 1805–1820. https://doi.org/https://doi.org/10.1002/jgre.20126

Cheek, L. C., Pieters, C. M., Boardman, J. W., Clark, R. N., Combe, J. P., Head, J. W., Isaacson, P. J., McCord, T. B., Moriarty, D., Nettles, J. W., Petro, N. E., Sunshine, J. M., & Taylor, L. A. (2011). Goldschmidt crater and the Moon's north polar region: Results from the Moon Mineralogy Mapper (M3). *Journal of Geophysical Research: Planets*, *116*(2). https://doi.org/10.1029/2010JE003702

Clark, R. N., King, T. V. V., & Gorelick, N. S. (1987). Automatic continuum analysis of reflectance spectra. JPL Proceedings of the 3rd Airborne Imaging Spectrometer Data Analysis Workshop.

Clark, R. N., Pieters, C. M., Green, R. O., Boardman, J. W., & Petro, N. E. (2011). Thermal removal from near-infrared imaging spectroscopy data of the Moon. *Journal of Geophysical Research E: Planets*, *116*(E6), 1–9. https://doi.org/10.1029/2010JE003751

Clark, J. D., van der Bogert, C. H., & Hiesinger, H. (2024). How old are lunar lobate scarps? 2. Distribution in space and time. *Earth and Planetary Science Letters*, *633*, 118636. https://doi.org/https://doi.org/10.1016/j.epsl.2024.118636

Cloutis, E. A., & Gaffey, M. J. (1991a). Spectral-compositional variations in the constituent minerals of mafic and ultramafic assemblages and remote sensing implications. *Earth, Moon, and Planets*, *53*(1), 11–53. https://doi.org/10.1007/BF00116217

Cloutis, E. A., & Gaffey, M. J. (1991b). Spectral-compositional variations in the constituent minerals of mafic and ultramafic assemblages and remote sensing implications. *Earth, Moon, and Planets*, *53*(1), 11–53. https://doi.org/10.1007/BF00116217

Cloutis, E. A., McCormack, K. A., Bell, J. F., Hendrix, A. R., Bailey, D. T., Craig, M. A., Mertzman, S. A., Robinson, M. S., & Riner, M. A. (2008). Ultraviolet spectral reflectance properties of common planetary minerals. *Icarus*, *197*(1), 321–347. https://doi.org/https://doi.org/10.1016/j.icarus.2008.04.018

Cohen, B. A., Snyder, G. A., Hall, C. M., Taylor, L. A., & Nazarov, M. A. (2001). Argon-40-argon-39 chronology and petrogenesis along the eastern limb of the Moon from Luna 16, 20 and 24 samples. 36(10), 1345–1366. https://doi.org/10.1111/j.1945-5100.2001.tb01829.x

Dasch, E. J., Shih, C.-Y., Bansal, B. M., Wiesmann, H., & Nyquist, L. E. (1987). Isotopic analysis of basaltic fragments from lunar breccia 14321: Chronology and petrogenesis of pre-Imbrium mare volcanism. *Geochimica et Cosmochimica Acta*, *51*(12), 3241–3254. https://doi.org/https://doi.org/10.1016/0016-7037(87)90132-3

Dombard, A. J., & Gillis, J. J. (2001). Testing the viability of topographic relaxation as a mechanism for the formation of lunar floor-fractured craters. *Journal of Geophysical Research: Planets*, *106*(E11), 27901–27909. https://doi.org/10.1029/2000JE001388

Donaldson Hanna, K. L., Cheek, L. C., Pieters, C. M., Mustard, J. F., Wyatt, M. B., & Greenhagen, B. T. (2012). Global Identifications of Crystalline Plagioclase Across the Lunar Surface Using M\^3 and Diviner Data. *43rd Annual Lunar and Planetary Science Conference, Lunar and Planetary Science Conference*, 1968.

Donaldson Hanna, K. L., Cheek, L. C., Pieters, C. M., Mustard, J. F., Greenhagen, B. T., Thomas, I. R., & Bowles, N. E. (2014). Global assessment of pure crystalline plagioclase across the Moon and implications for the evolution of the primary crust. *Journal of Geophysical Research: Planets*, *119*(7), 1516–1545. https://doi.org/https://doi.org/10.1002/2013JE004476

Elkins-Tanton, L. T., & Hager, B. H. (2005). Giant meteoroid impacts can cause volcanism. *Earth and Planetary Science Letters*, *239*(3), 219–232. https://doi.org/https://doi.org/10.1016/j.epsl.2005.07.029

Elkins-Tanton, L. T., Hager, B. H., & Grove, T. L. (2004). Magmatic effects of the lunar late heavy bombardment. *Earth and Planetary Science Letters*, *222*(1), 17–27. https://doi.org/10.1016/j.epsl.2004.02.017

Fassett, C. I., Head, J. W., Kadish, S. J., Mazarico, E., Neumann, G. A., Smith, D. E., & Zuber, M. T. (2012). Lunar impact basins: Stratigraphy, sequence and ages from superposed impact crater populations measured from Lunar Orbiter Laser Altimeter (LOLA) data. *Journal of Geophysical Research: Planets*, *117*(E12). https://doi.org/https://doi.org/10.1029/2011JE003951

Ghods, A., & Arkani-Hamed, J. (2007). Impact-induced convection as the main mechanism for formation of lunar mare basalts. *Journal of Geophysical Research: Planets*, *112*(E3). https://doi.org/https://doi.org/10.1029/2006JE002709

Green, R. O., Pieters, C., Mouroulis, P., Eastwood, M., Boardman, J., Glavich, T., Isaacson, P., Annadurai, M., Besse, S., Barr, D., Buratti, B., Cate, D., Chatterjee, A., Clark, R., Cheek, L., Combe, J., Dhingra, D., Essandoh, V., Geier, S., … Wilson, D. (2011). The Moon Mineralogy Mapper (M3) imaging spectrometer for lunar science: Instrument description, calibration, on-orbit measurements, science data calibration and on-orbit validation. *Journal of Geophysical Research: Planets*, *116*(E10). https://doi.org/https://doi.org/10.1029/2011JE003797

Gustafson, J. O., Bell III, J. F., Gaddis, L. R., Hawke, B. R., & Giguere, T. A. (2012). Characterization of previously unidentified lunar pyroclastic deposits using Lunar Reconnaissance Orbiter Camera data. *Journal of Geophysical Research: Planets*, *117*(E12). https://doi.org/https://doi.org/10.1029/2011JE003893

Gustafson, J. O., Gaddis, L. R., Bell, J. F., & Gustafson, J. A. (2020). An investigation of potential pyroclastic deposits on the southeast limb of the Moon. *Icarus*, *349*, 113828. https://doi.org/https://doi.org/10.1016/j.icarus.2020.113828

Hale, W. S., & Grieve, R. A. F. (1982). Volumetric analysis of complex lunar craters: Implications for basin ring formation. *Journal of Geophysical Research: Solid Earth*, *87*(S01), A65–A76. https://doi.org/https://doi.org/10.1029/JB087iS01p00A65

Hall, J. L., Solomon, S. C., & Head, J. W. (1981). Lunar floor-fractured craters: Evidence for viscous relaxation of crater topography. *Journal of Geophysical Research: Solid Earth*, *86*(B10), 9537–9552. https://doi.org/10.1029/jb086ib10p09537

Head, J. W., III (1974), Lunar dark-mantle deposits: Possible clues to the distribution of early mare deposits, *Proc. Lunar Sci. Conf.*, 5th, 207–222.

Head, J. W., & Wilson, L. (1992). Lunar mare volcanism: Stratigraphy, eruption conditions, and the evolution of secondary crusts. *Geochimica et Cosmochimica Acta*, *56*, 2155–2175.

Head, J. W., & Wilson, L. (2017). Generation, ascent and eruption of magma on the Moon: New insights into source depths, magma supply, intrusions and effusive/explosive eruptions (Part 2: Predicted emplacement processes and observations). *Icarus*, *283*, 176–223. https://doi.org/https://doi.org/10.1016/j.icarus.2016.05.031

Heiken, G. H., McKay, D. S., & Brown, R. W. (1974). Lunar deposits of possible pyroclastic origin. *Geochimica et Cosmochimica Acta*, *38*(11), 1703–1718. https://doi.org/https://doi.org/10.1016/0016-7037(74)90187-2

Johnson, B. C., Blair, D. M., Collins, G. S., Melosh, H. J., Freed, A. M., Taylor, G. J., Head, J. W., Wieczorek, M. A., Andrews-Hanna, J. C., Nimmo, F., Keane, J. T., Miljković, K., Soderblom, J. M., & Zuber, M. T. (2016). Formation of the Orientale lunar multiring basin. *Science*, *354*(6311), 441–444. https://doi.org/10.1126/science.aag0518

Jozwiak, L. M., Head, J. W., Neumann, G. A., & Wilson, L. (2015). The effect of evolving gas distribution on shallow lunar magmatic intrusion density: Implications for gravity anomalies. Lunar Planet. Sci. Conf., 46, abstract 1580

Jozwiak, L. M., Head, J. W., Zuber, M. T., Smith, D. E., & Neumann, G. A. (2012). Lunar floor-fractured craters: Classification, distribution, origin and implications for magmatism and shallow crustal structure. *Journal of Geophysical Research: Planets*, *117*(11). https://doi.org/10.1029/2012JE004134

Klima, R. L., PIETERS, C. M., & DYAR, M. D. (2008). Characterization of the 1.2 μm M1 pyroxene band: Extracting cooling history from near-IR spectra of pyroxenes and pyroxene-dominated rocks. *Meteoritics & Planetary Science*, *43*(10), 1591–1604. https://doi.org/https://doi.org/10.1111/j.1945-5100.2008.tb00631.x

Kramer, G. Y., Jolliff, B. L., & Neal, C. R. (2008). Distinguishing high-alumina mare basalts using Clementine UVVIS and Lunar Prospector GRS data: Mare Moscoviense and Mare Nectaris. *Journal of Geophysical Research: Planets*, *113*(E1). https://doi.org/https://doi.org/10.1029/2006JE002860

Laneuville, M., Wieczorek, M. A., Breuer, D., & Tosi, N. (2013). Asymmetric thermal evolution of the Moon. *Journal of Geophysical Research: Planets*, *118*(7), 1435–1452. https://doi.org/10.1002/jgre.20103

Lemelin, M., Lucey, P. G., Gaddis, L. R., Hare, T., & Ohtake, M. (2016). *Global Map Products from the Kaguya Multiband Imager at 512 ppd: Minerals, FeO, and OMAT*.

Lister, J. R. (1990). Buoyancy-driven fluid fracture: the effects of material toughness and of low-viscosity precursors. *Journal of Fluid Mechanics*, *210*, 263–280. https://doi.org/DOI: 10.1017/S0022112090001288

Lister, J. R., & Kerr, R. C. (1991). Fluid-mechanical models of crack propagation and their application to magma transport in dykes. *Journal of Geophysical Research: Solid Earth*, *96*(B6), 10049–10077. https://doi.org/https://doi.org/10.1029/91JB00600

McCord, T. B., Clark, R. N., Hawke, B. R., McFadden, L. A., Owensby, P. D., Pieters, C. M., & Adams, J. B. (1981). Moon: Near-infrared spectral reflectance, A first good look. *Journal of Geophysical Research: Solid Earth*, *86*(B11), 10883–10892. https://doi.org/https://doi.org/10.1029/JB086iB11p10883

Melosh, H. J., Impact Cratering: A Geologic Process, 245 pp., Oxford University Press, New York, 1989.

Michael, G. G., & Neukum, G. (2010). Planetary surface dating from crater size–frequency distribution measurements: Partial resurfacing events and statistical age uncertainty. *Earth and Planetary Science Letters*, *294*(3), 223–229. https://doi.org/https://doi.org/10.1016/j.epsl.2009.12.041

Morota, T., Haruyama, J., Honda, C., Ohtake, M., Yokota, Y., Kimura, J., Matsunaga, T., Ogawa, Y., Hirata, N., Demura, H., Iwasaki, A., Miyamoto, H., Nakamura, R., Takeda, H., Ishihara, Y., & Sasaki, S. (2009). Mare volcanism in the lunar farside Moscoviense region: Implication for lateral variation in magma production of the Moon. *Geophysical Research Letters*, *36*(21). https://doi.org/https://doi.org/10.1029/2009GL040472

Morota, T., Haruyama, J., Ohtake, M., Matsunaga, T., Honda, C., Yokota, Y., Kimura, J., Ogawa, Y., Hirata, N., Demura, H., Iwasaki, A., Sugihara, T., Saiki, K., Nakamura, R., Kobayashi, S., Ishihara, Y., Takeda, H., & Hiesinger, H. (2011). Timing and characteristics of the latest mare eruption on the Moon. *Earth and Planetary Science Letters*, *302*(3–4), 255–266. https://doi.org/10.1016/j.epsl.2010.12.028

Nahm, A. L., Öhman, T., & Kring, D. A. (2013). Normal faulting origin for the Cordillera and Outer Rook Rings of Orientale Basin, the Moon. *Journal of Geophysical Research: Planets*, *118*(2), 190–205. https://doi.org/https://doi.org/10.1002/jgre.20045

Nelson, D. M., Koeber, S. D., Daud, K., Robinson, M. S., Watters, T. R., Banks, M. E., & Williams, N. R. (2014). Mapping Lunar Maria Extents and Lobate Scarps Using LROC Image Products. *45th Annual Lunar and Planetary Science Conference, Lunar and Planetary Science Conference*, 2861.

Neukum, G., Ivanov, B. A., & Hartmann, W. K. (2001). Cratering Records in the Inner Solar System in Relation to the Lunar Reference System. *Space Science Reviews*, *96*(1), 55–86. https://doi.org/10.1023/A:1011989004263

Neumann, G. A., Zuber, M. T., Wieczorek, M. A., Head, J. W., Baker, D. M. H., Solomon, S. C., Smith, D. E., Lemoine, F. G., Mazarico, E., Sabaka, T. J., Goossens, S. J., Melosh, H. J., Phillips, R. J., Asmar, S. W., Konopliv, A. S., Williams, J. G., Sori, M. M., Soderblom, J. M., Miljkovic, K., … Kiefer, W. S. (2015). Planetary Science: Lunar impact basins revealed by gravity recovery and interior laboratory measurements. *Science Advances*, *1*(9). https://doi.org/10.1126/sciadv.1500852

Nyquist, L. E., Wooden, J. L., Shih, C.-Y., Wiesmann, H., & Bansal, B. M. (1981). Isotopic and REE studies of lunar basalt 12038: implications for petrogenesis of aluminous mare basalts. *Earth and Planetary Science Letters*, *55*(3), 335–355. https://doi.org/https://doi.org/10.1016/0012-821X(81)90162-X

Ohtake, M., Matsunaga, T., Haruyama, J., Yokota, Y., Morota, T., Honda, C., Ogawa, Y., Torii, M., Miyamoto, H., Arai, T., Hirata, N., Iwasaki, A., Nakamura, R., Hiroi, T., Sugihara, T., Takeda, H., Pieters, C., Saiki, K., & Josset, J.-L. (2009). The global distribution of pure anorthosite on the Moon. *Nature*, *461*, 236–240. https://doi.org/10.1038/nature08317

Ohtake, M., Pieters, C. M., Isaacson, P., Besse, S., Yokota, Y., Matsunaga, T., Boardman, J., Yamomoto, S., Haruyama, J., Staid, M., Mall, U., & Green, R. O. (2013). One Moon, Many Measurements 3: Spectral reflectance. *Icarus*, *226*(1), 364–374. https://doi.org/https://doi.org/10.1016/j.icarus.2013.05.010

Orgel, C., Michael, G., Fassett, C. I., van der Bogert, C. H., Riedel, C., Kneissl, T., & Hiesinger, H. (2018). Ancient Bombardment of the Inner Solar System: Reinvestigation of the "Fingerprints" of Different Impactor Populations on the Lunar Surface. *Journal of Geophysical Research: Planets*, *123*(3), 748–762. https://doi.org/https://doi.org/10.1002/2017JE005451

Panwar, N., Srivastava, N., Bhatt, M., & Bhardwaj, A. (2023). Compositional diversity in the Mare Marginis and Mare Smythii: An insight into the volcanism in the region. *Icarus*, *395*, 115496. https://doi.org/https://doi.org/10.1016/j.icarus.2023.115496

Panwar, N., Srivastava, N., Yadav, A., Bhatt, M., Wöhler, C., & Bhardwaj, A. (2025). Volcanism along the rings of the Crisium Basin on the Moon: Insights from M3 onboard Chandrayaan-1. *Icarus*, *438*, 116641. https://doi.org/https://doi.org/10.1016/j.icarus.2025.116641

Papanastassiou, D. A., & Wasserburg, G. J. (1971). RbSr ages of igneous rocks from the Apollo 14 mission and the age of the Fra Mauro formation. *Earth and Planetary Science Letters*, *12*(1), 36–48. https://doi.org/https://doi.org/10.1016/0012-821X(71)90052-5

Pieters, C. M., Boardman, J., Buratti, B., Chatterjee, A., Clark, R., Glavich, T., Green, R., Head, J., Isaacson, P., Malaret, E., McCord, T., Mustard, J., Petro, N., Runyon, C., Staid, M., Sunshine, J., Taylor, L., Tompkins, S., Varanasi, P., & White, M. (2009). The Moon Mineralogy Mapper ($M^3$) on Chandrayaan-1. *Current Science*, *96*(4), 500–505. http://www.jstor.org/stable/24105459

Pike, R. J. (1977). Size-dependence in the shape of fresh impact craters on the Moon. Impact and Explosion Cratering: Planetary and Terrestrial Implications, 489–509.

Pilkington, M., & Hildebrand, A. R. (2003). Transient and disruption cavity dimensions of complex terrestrial impact structures derived from magnetic data. *Geophysical Research Letters*, *30*(21). https://doi.org/https://doi.org/10.1029/2003GL018294

Robinson, M. S., Brylow, S. M., Tschimmel, M., Humm, D., Lawrence, S. J., Thomas, P. C., Denevi, B. W., Bowman-Cisneros, E., Zerr, J., Ravine, M. A., Caplinger, M. A., Ghaemi, F. T., Schaffner, J. A., Malin, M. C., Mahanti, P., Bartels, A., Anderson, J., Tran, T. N., Eliason, E. M., … Hiesinger, H. (2010). Lunar Reconnaissance Orbiter Camera (LROC) Instrument Overview. *Space Science Reviews*, *150*(1), 81–124. https://doi.org/10.1007/s11214-010-9634-2

Rosanova, C. E., Gaddis, L., Hare, T. M., Coombs, C., Hawke, B. R., & Robinson, M. S. (1998). Characterization of ``New'' Pyroclastic Deposits on the Moon Using Clementine Data. *Lunar and Planetary Science Conference, Lunar and Planetary Science Conference*, 1710.

Ruj, T., Komatsu, G., Kawai, K., Okuda, H., Xiao, Z., & Dhingra, D. (2022). Recent boulder falls within the Finsen crater on the lunar far side: An assessment of the possible triggering rationale. *Icarus*, *377*, 114904. https://doi.org/https://doi.org/10.1016/j.icarus.2022.114904

Schultz, P. H. (1976), Floor-fractured lunar craters, Moon, 15, 241–273. doi:10.1007/BF00562240.

Schultz, van der Bogert, C., & Pieters, C. (1998). *The Possible Generation of Friction Melts at the Lunar Crater, Buys-Ballot*.

Senthil Kumar, P., Sruthi, U., Krishna, N., Lakshmi, K. J. P., Menon, R., Amitabh, Gopala Krishna, B., Kring, D. A., Head, J. W., Goswami, J. N., & Kiran Kumar, A. S. (2016). Recent shallow moonquake and impact-triggered boulder falls on the Moon: New insights from the Schrödinger basin. *Journal of Geophysical Research: Planets*, *121*(2), 147–179. https://doi.org/https://doi.org/10.1002/2015JE004850

Speyerer, E. J., Robinson, M. S., Denevi, B. W., & LROC Science Team. (2011). Lunar Reconnaissance Orbiter Camera Global Morphological Map of the Moon. *42nd Annual Lunar and Planetary Science Conference, Lunar and Planetary Science Conference*, 2387.

Spudis, P. D., Gillis, J. J., & Reisse, R. A. (1994). Ancient Multiring Basins on the Moon Revealed by Clementine Laser Altimetry. *Science*, *266*(5192), 1848–1851. https://doi.org/10.1126/science.266.5192.1848

Srivastava, N., & Gupta, R. ~P. (2013). Spatial Distribution of Spinel in the Orientale Basin: New Insights from M\^3 Data. *44th Annual Lunar and Planetary Science Conference, Lunar and Planetary Science Conference*, 1509.

Stuart-Alexander, D. E. (1978). Geologic map of the central far side of the Moon. In *IMAP*. https://doi.org/10.3133/i1047

Taguchi, M., Morota, T., & Kato, S. (2017). Lateral heterogeneity of lunar volcanic activity according to volumes of mare basalts in the farside basins. *Journal of Geophysical Research: Planets*, *122*(7), 1505–1521. https://doi.org/https://doi.org/10.1002/2016JE005246

Taisne, B., & Tait, S. (2009). Eruption versus intrusion? Arrest of propagation of constant volume, buoyant, liquid-filled cracks in an elastic, brittle host. *Journal of Geophysical Research (Solid Earth)*, *114*(B6), B06202. https://doi.org/10.1029/2009JB006297

Taisne, B., Tait, S., & Jaupart, C. (2011). Conditions for the arrest of a vertical propagating dyke. *Bulletin of Volcanology*, *73*(2), 191–204. https://doi.org/10.1007/s00445-010-0440-1

Thorey, C., & Michaut, C. (2014). A model for the dynamics of crater-centered intrusion: Application to lunar floor-fractured craters. *Journal of Geophysical Research: Planets*, *119*(1), 286–312. https://doi.org/10.1002/2013JE004467

van der Bogert, C. H., Clark, J. D., Hiesinger, H., Banks, M. E., Watters, T. R., & Robinson, M. S. (2018). How old are lunar lobate scarps? 1. Seismic resetting of crater size-frequency distributions. *Icarus*, *306*, 225–242. https://doi.org/https://doi.org/10.1016/j.icarus.2018.01.019

Vijayan, S., Sharini, K. S., Kimi, K. B., Tuhi, S., Harish, Sahoo, R., Watters, T. R., & Bhardwaj, A. (2025). Recent boulder falls on the Moon. *Icarus*, *438*, 116627. https://doi.org/https://doi.org/10.1016/j.icarus.2025.116627

Whitten, J., Head, J. W., Staid, M., Pieters, C. M., Mustard, J., Clark, R., Nettles, J., Klima, R. L., & Taylor, L. (2011). Lunar mare deposits associated with the Orientale impact basin: New insights into mineralogy, history, mode of emplacement, and relation to Orientale Basin evolution from Moon Mineralogy Mapper (M3) data from Chandrayaan-1. *Journal of Geophysical Research: Planets*, *116*(4). https://doi.org/10.1029/2010JE003736

Wichman, R. W., & Schultz, P. H. (1995). Floor-fractured craters in Mare Smythii and west of Oceanus Procellarum: Implications of crater modification by viscous relaxation and igneous intrusion models. *Journal of Geophysical Research: Planets*, *100*(E10), 21201–21218. https://doi.org/https://doi.org/10.1029/95JE02297

Wieczorek, M. A., Neumann, G. A., Nimmo, F., Kiefer, W. S., Taylor, G. J., Melosh, H. J., Phillips, R. J., Solomon, S. C., Andrews-Hanna, J. C., Asmar, S. W., Konopliv, A. S., Lemoine, F. G., Smith, D. E., Watkins, M. M., Williams, J. G., & Zuber, M. T. (2013). The Crust of the Moon as Seen by GRAIL. *Science*, *339*(6120), 671–675. https://doi.org/10.1126/science.1231530

Wieczorek, M. A., Zuber, M. T., & Phillips, R. J. (2001). The role of magma buoyancy on the eruption of lunar basalts. *Earth and Planetary Science Letters*, *185*(1), 71–83. https://doi.org/https://doi.org/10.1016/S0012-821X(00)00355-1

Wilhelms, D. E., McCauley, J. F., & Trask, N. J. (1987). The geologic history of the Moon. In *Professional Paper*. https://doi.org/10.3133/pp1348

Wilson, L., & Head, J. W. (2017). Generation, ascent and eruption of magma on the Moon: New insights into source depths, magma supply, intrusions and effusive/explosive eruptions (Part 1: Theory). *Icarus*, *283*, 146–175. https://doi.org/https://doi.org/10.1016/j.icarus.2015.12.039

Yamamoto, S., Nakamura, R., Matsunaga, T., Ogawa, Y., Ishihara, Y., Morota, T., Hirata, N., Ohtake, M., Hiroi, T., Yokota, Y., & Haruyama, J. (2012). Massive layer of pure anorthosite on the Moon. *Geophysical Research Letters*, *39*(13). https://doi.org/https://doi.org/10.1029/2012GL052098

Zhang, X., & Cloutis, E. (2021a). Near-infrared Spectra of Lunar Ferrous Mineral Mixtures. *Earth and Space Science*, *8*(4). https://doi.org/10.1029/2020EA001153

Zhang, X., & Cloutis, E. (2021b). Variations in the Near-Infrared Spectral Properties of Ferrous Mineral Mixtures With Different Relative Abundances. *Earth and Space Science*, *8*(9). https://doi.org/10.1029/2021EA001636

Zhang, L., Zhang, X., Yang, M., Xiao, X., Qiu, D., Yan, J., Xiao, L., & Huang, J. (2023). New maps of major oxides and Mg # of the lunar surface from additional geochemical data of Chang'E-5 samples and KAGUYA multiband imager data. *Icarus*, *397*, 115505. https://doi.org/https://doi.org/10.1016/j.icarus.2023.115505